\documentstyle[12pt,epsfig,epstopdf,subfigure,color,amsfonts,
amssymb,inputenc,graphicx,amsmath,amsxtra,amstext,slashed,float]{article}
\newcommand{\be}{\begin{equation}}
\newcommand{\ee}{\end{equation}}
\newcommand{\beq}{\begin{equation}}
\newcommand{\eeq}{\end{equation}}
\newcommand{\ba}{\begin{eqnarray}}
\newcommand{\ea}{\end{eqnarray}}
\newcommand{\bea}{\begin{eqnarray}}
\newcommand{\eea}{\end{eqnarray}}

\begin{document}
\baselineskip=15.5pt \pagestyle{plain} \setcounter{page}{1}
%
%--------+---------+---------+---------+---------+---------+---------+
%--------+---------+---------+---------+---------+---------+---------+
\begin{titlepage}

\vskip 0.8cm

\begin{center}
%
%
%
%--------+---------+---------+---------+---------+---------+---------+
%--------+---------+---------+---------+---------+---------+---------+
%\begin{titlepage}
%
%\vskip 0.8cm
%
%\begin{center}

{\Large \bf  Spin-1/2 fermionic operators of ${\cal {N}}=4$ SYM theory and DIS from type IIB supergravity}

\vskip 1.cm

{\large {{\bf David Jorrin}{\footnote{\tt jorrin@fisica.unlp.edu.ar}}, {\bf Gustavo Michalski}{\footnote{\tt michalski@fisica.unlp.edu.ar}}, {\bf and  Martin
Schvellinger}{\footnote{\tt martin@fisica.unlp.edu.ar}}}}

\vskip 1.cm

{\it Instituto de F\'{\i}sica La Plata-UNLP-CONICET. \\
Boulevard 113 e 63 y 64, (1900) La Plata, Buenos Aires, Argentina \\
and \\
Departamento de F\'{\i}sica, Facultad de Ciencias Exactas,
Universidad Nacional de La Plata. \\
Calle 49 y 115, C.C. 67, (1900) La Plata, Buenos Aires, Argentina.}

\vspace{1.cm}

{\bf Abstract}

\end{center}

\vspace{1.cm}

We study deep inelastic scattering (DIS) of charged leptons from polarised spin-1/2 hadrons in terms of the gauge/gravity duality. We calculate the structure functions related to spin-1/2 fermionic operators of ${\cal {N}}=4$ SYM theory in the planar limit and at strong coupling. Specifically, we focus on the twist-3 operator which is dual to a certain dilatino mode and gives the leading contribution to the hadronic tensor. We consider the Bjorken variable within the parametric range $\lambda^{-1/2}_{SYM} \ll x < 1$ where the supergravity dual description holds. From dimensional reduction of type IIB supergravity on the five-sphere, we derive the interaction terms involving two dilatini and a massless vector field mode. This vector field is a linear combination of certain components of the graviton and the four-form potential. The angular integrals on the five-sphere lead to selection rules for the interactions with important consequences on the dynamics. This implies the existence of new terms contributing to the structure functions that we explicitly calculate. The leading contribution comes from the Pauli term, followed by the contributions from the new terms we found.  

\noindent

\end{titlepage}

\newpage

{\small \tableofcontents}

\newpage

%%%%%%%%%%%%%%%%%%%%%%%%%%%%%%%%%%%%%%%%%%%%%%%%%%%%%%%%%%%%%%%%%%%%%%%%

%---------------------------------------------------------------------
%
\section{Introduction}
%
%---------------------------------------------------------------------

In this work we calculate the hadronic tensor related to certain spin-1/2 fermionic operators of the strongly coupled ${\cal {N}}=4$ supersymmetric Yang-Mills (SYM) theory with gauge group $SU(N)$, in the large-$N$ limit, considering the range $\lambda_{SYM}^{-1/2} \ll x < 1$, where $x$ is the Bjorken parameter, while the 't Hooft coupling is defined as $\lambda_{SYM} \equiv g_{YM}^2 N$. This is carried out within the framework of the gauge/gravity duality, {\it i.e.} in terms of type IIB supergravity. The physical process corresponds to the polarised deep inelastic scattering (DIS) of charged leptons off spin-1/2 hadrons. We assume the hadron to be represented by a dilatino field mode of type IIB supergravity on AdS$_5 \times S^5$. We focus on the leading contribution to the hadronic tensor in the high energy limit ($\Lambda^2 \ll q^2$) which in the above conditions is given by the twist-3 fermionic operator ${\cal {O}}^{(6)}_{k=0}$ defined below.

In a pioneering work Polchinski and Strassler \cite{Polchinski:2002jw} have used the gauge/gravity duality to calculate the structure functions of glueballs as well as spin-1/2 fermions. In order to induce confinement in the gauge field theory they introduce an IR  cutoff in the AdS$_5$ spacetime. They distinguish three different dynamical regimes in terms of the Bjorken parameter: a) the case $\lambda_{SYM}^{-1/2} \ll x < 1$, where the dual description is given in terms of type IIB supergravity; b) the case when $\exp{(-\lambda_{SYM}^{1/2})} \ll x \ll \lambda_{SYM}^{-1/2}$, for which flat-space type IIB superstring theory scattering amplitudes are convoluted with the background metric AdS$_5 \times S^5$ within the so-called {\it ultra-local} approximation; and finally c) the regime $x \sim\exp{(-\lambda_{SYM}^{1/2})}$ where that approximation breaks down and therefore the holographic Pomeron techniques are useful to calculate the structure functions. Specifically, in \cite{Polchinski:2002jw} glueballs have been studied in the three parametric regimes. On the other hand, for the spin-1/2 fermions they only studied the supergravity regime by considering only the minimal coupling, which as we show in the present work, does not lead to the complete answer in terms of a consistent spontaneous compactification of type IIB supergravity on the five-sphere. Gao and Xiao \cite{Gao:2009ze} have studied DIS and elastic scattering on a polarised spin-1/2 hadron in similar terms as in \cite{Polchinski:2002jw}, {\it i.e.}  considering the virtual photon represented by a graviphoton, but not including the fluctuations of the four-form potential as we will describe below. Later on Gao and Mou \cite{Gao:2010qk} have considered some effects of introducing {\it by hand} a Pauli term in an effective five-dimensional action. They also have shown how to obtain the Pauli term from a 6-dimensional model, however they have not derived this term from a consistent dimensional reduction of type IIB supergravity on $S^5$. Moreover, in \cite{Gao:2009ze} they associate the minimal coupling with the calculation of the polarised proton structure functions. On the other hand, in \cite{Gao:2010qk} they focus on the polarised neutron-like structure functions where only the Pauli term appears. From a purely five-dimensional bottom-up perspective, in principle, there would not be an apparent restriction to switch on/off the minimal interaction and Pauli interaction terms as done in \cite{Gao:2009ze} and \cite{Gao:2010qk}. 

The first task we carry out in the present work is a first-principles derivation of the five-dimensional interaction of two dilatini with a vector field from type IIB supergravity. This is essential since it makes  possible for us to calculate the relative constants between the Pauli and the minimal interaction terms. Then, from our present calculation based on a consistent compactification of type IIB supergravity we show that all dilatino modes become charged under the isometry group of the five-sphere. Thus, both the Pauli interaction and minimal interaction terms contribute in all cases, {\it i.e.} from a top-down holographic dual calculation it is not possible to turn on/off by hand any of those terms. In this way, it exposes certain important limitations related to the bottom-up approach.

We would like to emphasize that the consistent approach in the present context implies that the spin-1/2 hadron, holographically represented by a dilatino field mode, interacts with a virtual photon, which is represented by a supergravity massless vector field mode. This vector field is precisely given by a linear combination of the off-diagonal graviton components and certain components of the four-form potential of the ten-dimensional type IIB supergravity theory. The consistent reduction on $S^5$ of the spin-1/2 fermionic sector including interactions is in fact a non-trivial task. As already mentioned we have done it specifically for the cubic interaction terms involving two dilatini and a massless vector field. This is very interesting because it permits to unveil very important new effects for the DIS problem of the ${\cal {N}}=4$ SYM theory that we have found and introduce in the following sections. In addition, for the small-$x$ regime a heuristic approach has been developed in \cite{Hatta:2009ra}, while a first-principles derivation of all structure functions for the spin-1/2 fermions has been developed in \cite{Kovensky:2018xxa} where the small-$x$ regime has been investigated from closed string scattering amplitudes in type IIB superstring theory for polarised DIS. A very important remaining aspect is to develop a fully consistent holographic dual description of polarised DIS of charged leptons from spin-1/2 hadrons in the whole range of the Bjorken parameter. Therefore, in order to carry out this programme we need to investigate the structure functions of the hadronic tensor in the $\lambda_{SYM}^{-1/2} \ll x < 1$ range, in terms of a fully consistent treatment from type IIB supergravity. This is what we carry out in the present work.

Specifically, we consider the contributions to the hadronic tensor given by the operators: ${\cal {O}}_k^{(6)} \sim$tr$(F_+ \lambda_{{\cal {N}}=4} X^k)$ and ${\cal {O}}_k^{(13)} \sim$ tr$(F_+^2 \overline{\lambda}_{{\cal {N}}=4} X^k)$ with $k \ge 0$, where the trace is taken over the adjoint representation of $SU(N)$ \cite{DHoker:2002nbb}. $\lambda_{{\cal {N}}=4}$ represents left Weyl fermions\footnote{Notice that $\lambda_{{\cal {N}}=4}$ represents the four-dimensional ${{\cal {N}}=4}$ SYM Weyl fermions while $\hat{\lambda}$ denotes the ten-dimensional dilatino field of type IIB supergravity.} while $X$ are real scalars. They all belong to the gauge supermultiplet of ${\cal {N}}=4$ SYM theory. $F_+$ is the self-dual two-form field strength. The ${\cal {O}}_k^{(6)}$ operators belong to the $SU(4)_R$ irreducible representations {\bf 4$^*$}, {\bf 20$^*$}, {\bf 60$^*$}, and so on, with $k=0, 1, 2,$ etc., being their $SO(1,3) \times SO(1,1) \times SU(4)_R$ quantum numbers $(1/2,0)$, $\Delta=k+\frac{7}{2}$, and $(1, k, 0)$, respectively. On the other hand, for ${\cal {O}}_k^{(13)}$ operators the corresponding $SU(4)_R$ irreducible representations are {\bf 4}, {\bf 20}, {\bf 60}, and so on for the $k=0, 1, 2,$ etc. Their $SO(1,3) \times SO(1,1) \times SU(4)_R$ quantum numbers are $(0, 1/2)$, $\Delta=k+\frac{11}{2}$, and $(0, k, 1)$, respectively. In order to connect these operators with their corresponding dual dilatino field modes, recall that after dimensional reduction on $S^5$ there are two towers of dilatino Kaluza-Klein modes $\lambda_k^\pm$ \cite{Kim:1985ez}. The tower of five-dimensional dilatino modes $\lambda_k^-$ with (five-dimensional) masses $m_k^-=-k-\frac{3}{2}$ are dual to the ${\cal {O}}_k^{(6)}$ operators. In addition, the tower of $\lambda_k^+$ modes with masses $m_k^+=k+\frac{7}{2}$ are dual to the ${\cal {O}}_k^{(13)}$ operators.

The dimensional reduction of type IIB supergravity on the five-sphere including the dilatino interactions (in ten dimensions) leads to a five-dimensional effective supergravity action which includes interaction terms beyond the minimal coupling for the dilatino. We would like to emphasize that we carry out a first-principles derivation from type IIB supergravity, which is distinct from other previous papers focusing on the DIS structure functions of spin-1/2 hadrons. Also, as we shall explain below this derivation of the cubic interactions with two dilatini and a vector field mode has not been derived previously, at least in the way we present it. The angular integrals lead to very interesting selection rules for the interactions with important consequences on the dynamics for both supergravity and ${\cal{N}}=4$ SYM theory. These imply the existence of new terms in the calculation of the structure functions of the spin-1/2 hadrons in comparison with references \cite{Polchinski:2002jw,Gao:2009ze,Gao:2010qk}. We have explicitly calculated the contributions of each type of interaction to the structure functions, and obtained that the contributions from the minimal coupling given in \cite{Polchinski:2002jw} are very small as shown in figures 3, 4, 5 and 6 and table 1 of our present work, while the Pauli term renders very substantial contributions to them. Although the Pauli term was considered in \cite{Gao:2010qk} we have obtained the first complete calculation of the relative contributions of both terms. Moreover, we have found that the new terms, those given by the selection rules which we have obtained in this work, as well as the ones obtained from the Feynman diagrams of forward Compton scattering with both the minimal coupling vertex and the Pauli vertex in the $s$-channel, provide additional contributions to the structure functions which are very important in comparison with the minimal coupling contribution. Also, from type IIB supergravity we derive relations between different structure functions.

In order to derive the interactions from type IIB supergravity we  follow a similar procedure as previously proposed in references \cite{Lee:1998bxa} and \cite{Arutyunov:1999en}, where it has been analysed cubic interaction terms with $s^I$ scalars \cite{Lee:1998bxa} and  $s^I$, $t^I$ and $\phi^I$ scalars \cite{Arutyunov:1999en}, in order to obtain three-point correlation functions of a proper extension of chiral primary operators \cite{Arutyunov:1999en}. In our present work the crucial difference is that we now consider cubic interactions of two spin-1/2 fermions coupled to a vector field mode\footnote{Note that in reference \cite{Uruchurtu:2007kq} interactions of two dilatini and a fluctuation of the axio-dilaton have been derived from type IIB supergravity. This is totally different in comparison with our present work, where the relevant coupling of the two dilatini involves a massless vector field.}. Due to the lack of a simple covariant action for type IIB supergravity due to the self-duality condition, it is proposed to work with the covariant equations of motion (EOM) in order to firstly obtain the quadratic terms and from them one can construct an effective interaction Lagrangian. Besides, another possibility could be to consider the covariant action proposed in reference \cite{DallAgata:1998ahf}, however in this case the calculation turns out to be very complicated due to the auxiliary fields that it contains. It is interesting to emphasize that in order to calculate the normalisation constant between the two- and three-point correlation functions the authors of reference \cite{Lee:1998bxa} have done a comparison of their results with those of reference \cite{DallAgata:1998ahf}. In our case, the unknown normalisation constants of the structure functions of the holographic hadrons are just overall factors, which can be written in terms of a single overall constant. Thus, we anticipate that the final result of the complete calculation of the hadronic tensor will just have a single normalisation constant ($a_0$ defined in Section 4.2), {\it i.e.} there is only one free parameter which is the same for all the structure functions.

The application of the AdS/CFT duality to deep inelastic scattering has been studied in a number of interesting papers. For instance, investigations including the eikonal approach have been done in \cite{Brower:2007xg,Brower:2007qh,Cornalba:2006xm,Cornalba:2007zb,Hatta:2007he,Nishio:2011xz,Costa:2012fw,Watanabe:2012uc,Costa:2013uia,Nally:2017nsp}. Also, other different aspects of deep inelastic scattering have been studied in terms of the AdS/CFT correspondence in \cite{BallonBayona:2007qr,BallonBayona:2008zi,BallonBayona:2009uy,BallonBayona:2010ae,Bayona:2011xj,Ballon-Bayona:2017vlm}. In reference \cite{Koile:2011aa} it has been studied DIS of charged leptons 
from holographic scalar and vector mesons for the D3D7, D4D8 anti-D8, and D4D6 anti-D6-brane models for a single flavor, while in \cite{Koile:2013hba} it was considered the multi-flavored case in terms of both type IIA and type IIB 
supergravities. Scalar and vector mesons structure functions in the $x << {\lambda_{SYM}}^{-1/2}$ regime using superstring theory have been calculated in \cite{Koile:2014vca} and \cite{Koile:2015qsa}. Moreover, $1/N^2$ corrections to deep inelastic scattering from the gauge/gravity duality have been derived for glueballs in \cite{Jorrin:2016rbx}, while in \cite{Kovensky:2016ryy} the $1/N$ expansion from the D3D7-brane system has been obtained for scalar mesons. Also, $1/N$ corrections to $F_1$ and $F_2$ structure functions of vector mesons from type IIB supergravity have been obtained in \cite{Kovensky:2018gif}. In these cases the interesting result is that $1/N$ corrections fit substantially better the corresponding lattice QCD results for the pion and the $\rho$ meson in comparison with previous results in the planar limit. Moreover, in \cite{Kovensky:2017oqs} the role of the chiral anomaly and the Chern-Simons term in the structure of glueballs has been investigated. Another compelling result is presented in
\cite{Kovensky:2018xxa} where DIS of charged leptons from polarised spin-$1/2$ hadrons has been investigated at small $x$ from type IIB superstring theory, obtaining the antisymmetric structure function $g_1$, which has been fitted to experimental data from COMPASS Collaboration \cite{Aghasyan:2017vck} for the corresponding function $g_1^p$ of the proton, with a $\chi^2$ per degree of freedom of 1.074. This encourages to continue exploring high energy scattering processes in particle physics in terms of the AdS/CFT duality.

The work is organized as follows. In Section 2 we obtain the cubic interaction terms of two dilatini and a vector field from type IIB supergravity. We firstly derive the EOM of the dilatino field coupled to a vector field mode $B_a^l$ with $l \geq 1$, which is a linear combination of the off-diagonal components of the graviton and the AdS$_5$ vector components of the four-form potential. Then, we consider the special case of $l=1$, {\it i.e.} the massless vector field mode $B_a^1$, and obtain the explicit form of the contribution of all the corresponding cubic interaction terms involving two dilatini and a massless vector field to the five-dimensional supergravity action. In Section 3 we give some definitions of the hadronic tensor. In Section 4 calculate explicitly all the corresponding structure functions of the hadronic tensor using the holographic dual prescription, focusing on the case of the twist-3 spin-1/2 fermionic operator. In Section 5 we introduce our results corresponding to polarised structure functions associated with the twist-3 spin-1/2 fermionic operator. In Section 6 we write the discussion and the conclusions. Some details of the calculations are shown in two appendices.

%-----------------------------------------------------------------------
%
\section{Cubic interaction terms of two dilatini and a vector field from type IIB supergravity}
%
%-----------------------------------------------------------------------

The five-dimensional IIB supergravity compactification on AdS$_5 \times S^5$ has been carried out by Kim, Romans and van Nieuwenhuizen using harmonic analysis on the five-sphere \cite{Kim:1985ez}, obtaining the mass spectrum for all fields. 
In this section we derive the five-dimensional cubic interaction terms between two dilatini and a vector field from type IIB supergravity. This vector field $B_a$ is a linear combination of off-diagonal graviton fluctuations of the form $h_{a\alpha}$, and the four-form potential fluctuations with one index on AdS$_5$ and three indices on $S^5$, denoted by $a_{a\alpha\beta\gamma}$, being both ten-dimensional fields of type IIB supergravity. The vector field modes can be massless or massive, and we discuss both situations in general terms. As anticipated in the introduction we carry out a procedure which is somehow anologous to the derivation of the cubic interaction terms involving two scalars of the type $s^I$ and one scalar of the type $t^I$ or $\phi^I$  developed in references \cite{Lee:1998bxa,Arutyunov:1999en}\footnote{Recall that the scalars $s^I$ and $t^I$ are mixtures of the four-form potential fluctuations $a_{\alpha\beta\gamma\delta}$ on $S^5$ with the trace of the graviton on $S^5$ ($h_\alpha^{\,\,\alpha}$), while the scalar $\phi^I$ is given by the graviton fluctuations on $S^5$ ($h_{\alpha\beta}$).}, in the sense that we begin with the linearised EOMs of the dilatino and the field $B_a$ in type IIB supergravity and add fluctuations to them. To our knowledge the interactions between dilatini and the vector field modes $B_a$ that we derive in this work have not been considered in previous literature.

Let us recall that the bosonic field content of type IIB supergravity includes the graviton, a complex scalar, a complex two-index antisymmetric tensor, and a real four-index antisymmetric tensor $A_{PQRS}$ that we call the four-form potential. $M,N,P, \dots=0,\cdots, 9$ denote ten-dimensional curved indices. On the other hand, the fermionic sector contains a chiral complex gravitino and a chiral complex spin-1/2 fermion (the dilatino $\hat{\lambda}$) of opposite chirality.

\subsection{Derivation of the dilatino EOM coupled to a vector field $B_a^l$}

In order to derive the cubic interaction terms with two dilatini and a vector field let us consider the covariant EOMs of type IIB supergravity fields. Then, we can add second order corrections in the supergravity fields to the linearised contributions. This method has been used by Lee, Minwalla, Rangamani and Seiberg in \cite{Lee:1998bxa} considering cubic interaction terms involving $s^I$ scalars. Also, Arutyunov and Frolov used these ideas to obtain cubic and quartic Lagrangians \cite{Arutyunov:1999en,Arutyunov:1999fb}. In our case we must derive interactions between two dilatini $\hat{\lambda}$ and a vector field mode $B_a^l$. Specifically, for the DIS calculation we consider we need the massless mode of this vector field $B_a^1$, {\it i.e.} $l=1$.

The type IIB supergravity EOMs for the ten-dimensional metric and the five-form field strength are \cite{Kim:1985ez} \footnote{There is a minus sign of difference in comparison with equation (2.1) of Kim, Romans and van Nieuwenhuizen (KRvN) \cite{Kim:1985ez}. It comes from the definition of the Ricci tensor which we define as $R_{MN}=R^K_{\,\,\,\,MNK}$ as in \cite{Lee:1998bxa}, while in \cite{Kim:1985ez} it is defined as $R_{MN}^{(KRvN)}=R^K_{\,\,\,\,MKN}$. Therefore, our Ricci tensor has an overall minus sign with respect to the KRvN's one $R_{MN}=-R_{MN}^{(KRvN)}$.}
\begin{eqnarray}
R_{MN}=\frac{1}{3!}F_{MPQRS}F_{N}^{\,\,\,\,PQRS}, \ \ \ \ \ \ \ \ \ F_5=*F_5 \, 
,
\end{eqnarray}
where $M, N, O, \dots $ are curved indices in ten dimensions. The second equation is the self-duality condition for $F_5$. The symbol $*$ stands for the ten-dimensional Hodge dual operator. The five-form field strength is given by
\begin{equation}
F_{MPQRS}=5\partial_{[M}A_{PQRS]}=\partial_M A_{PQRS}+4 \  \text{terms} \, .
\end{equation}
A solution of the above equations is given by the AdS$_5\times S^5$ background with a constant five-form field strength and a constant dilaton field, while the rest of supergravity fields are zero,
\begin{eqnarray}
ds^2&=&g^0_{M N} dx^M dx^N= \frac{dz^2 +  \eta_{\mu\nu} dx^\mu dx^\nu}{z^2}+d\Omega_5^2 \, , \\
F^{0}_{abcde}&=&\epsilon_{abcde}, \ \ \ \ \ \ \ \ \  F^{0}_{\alpha \beta \gamma \delta \epsilon}= \epsilon_{\alpha \beta \gamma \delta \epsilon} \, , 
\end{eqnarray}
where $a, \cdots e$ and $\alpha, \cdots, \epsilon$ are
AdS$_5$ and $S^5$ curved indices, respectively. We use the mostly plus four-dimensional Minkowski metric $\eta_{\mu\nu} =$ diag$(-1,+1,+1,+1)$. We have set to one the radius of $S^5$ as well as the scale of the AdS$_5$ space. Now, we consider the fluctuations of the ten-dimensional type IIB supergravity fields which can be written as follows
\begin{eqnarray}
g_{MN}&=&g_{MN}^0+h_{MN} \, , \\
A_{MNOP}&=&A^{0}_{MNOP}+a_{MNOP} \, , \ \ \ \ F_{MNOPR}=F^0_{MNOPR} + f_{MNOPR} \, ,
\end{eqnarray}
where the label $0$ indicates the background fields solution.

The EOMs for the bosonic and fermionic fields linearised in fluctuations of the fields on the AdS$_5 \times S^5$ background have been obtained in reference \cite{Kim:1985ez}. Bosonic and fermionic fluctuations have been expanded in scalar, vector and spinor spherical harmonics on $S^5$ respectively, obtaining the bosonic and fermionic mass spectra. In the present work we are interested in the interactions of the dilatino with the massless mode of the vector field $B_a$. Since this field is obtained from the off-diagonal metric perturbations $h_{a\alpha}$ and the four-form potential perturbations $a_{a\alpha\beta\gamma}$, we focus on their EOMs. In order to fix the redundancies coming from the diffeomorphism  invariance one imposes the de Donder gauge\footnote{$\nabla^{\alpha}$ denotes the covariant derivative.}
\begin{eqnarray}
   \nabla^{\alpha}h_{a \alpha}=\nabla^{\alpha}h_{(\alpha \beta)}=0 \, , \ \ \ \ \ \  h_{(\alpha \beta)}=h_{\alpha \beta}-\frac{1}{5}g_{\alpha \beta}h_{\gamma}^{\gamma} \, , \\
   \nabla^{\alpha}a_{\alpha \beta \gamma \delta}=\nabla^{\alpha}a_{\alpha \beta \gamma a}=\nabla^{\alpha}a_{\alpha \beta a b}=\nabla^{\alpha}a_{\alpha a b c}=0 \, ,
\end{eqnarray}
where $(\alpha\beta)$ indicates symmetrization and traceless.  This allows to eliminate certain fluctuations, thus simplifying the EOMs. After imposing the de Donder gauge on the off-diagonal metric fluctuations, they can be expanded in vector spherical harmonics on $S^5$ \cite{Kim:1985ez}\footnote{$I_5$ represents the five indices ($l_5, l_4, l_3, l_2, l_1$) of the spherical harmonics on $S^5$. As shown in equation (\ref{mov-eq-arm}) there is a mass degeneracy, {\it i.e.} it only depends on $l_5$. Thus, in order to make the notation simpler we can set $l_5 \equiv l$ and drop the indices $l_1, l_2, l_3, l_4$. Through this work we use both $I_5$ (meaning the above five indices) and $l=l_5$.}
\begin{eqnarray}
h_{a \alpha}(x,y)&=&\sum_{I_5} A_{a}^{I_5}(x) Y_{\alpha}^{I_5}(y)\\
(\Box-4)Y_{\alpha}^l(y)&=&-(l+1)(l+3)Y_{\alpha}^l(y) \, ,
\label{mov-eq-arm}
\end{eqnarray}
where $l=1, 2, \dots$. The coordinates $x$ are on AdS$_5$ space while the $y$'s are on $S^5$.

The fluctuations of the four-form potential which are relevant for us are
\begin{eqnarray}
a_{a \alpha \beta \gamma }= \sum_{I_5} \Phi^{I_5}_a(x) \,
\epsilon_{\alpha \beta \gamma \delta \epsilon} \, \nabla^{\delta}Y^{I_5\epsilon}(y) \, , \ \ \ \ \ \ 
a_{\alpha a b c}=\sum_{I_5} b^{I_5}_{a b c}(x) \, Y^{I_5}_{\alpha}(y) \, . \label{fluc-5-form}
\end{eqnarray}
The second fluctuation can be rewritten as
\begin{eqnarray}
  a_{\alpha a b c}=\sum_{I_5} \epsilon_{abcde} \nabla^d\Phi^{I_5 e}(x) Y^{I_5}_{\alpha}(y) \, .
\end{eqnarray}
The corresponding EOMs to the fluctuations $A_{a}^{I_5}(x)$ and $\Phi^{I_5 e}(x)$ are coupled Maxwell-Einstein equations. After diagonalizing these equations the eigenvectors are the vector modes $B^l_a$ and $C^l_a$, while their corresponding eigenvalues are $M^2_{B, l}$ and $M^2_{C, l}$ for $l \geq 1$,
\begin{eqnarray}
B_a^l&=&A_a^l-4(l+3)\Phi_a^l \, , \ \ \ \ \  M^2_{B, l}=(l^2-1) \, , \label{B-modes} \\
C_a^l&=& A_a^l+4(l+1)\Phi_a^l \, , \ \ \ \ \ M^2_{C, l}=(l+3)(l+5) \, . \label{C-modes}
\end{eqnarray}
The corresponding irreducible representations of the $SU(4) \sim SO(6)$ group are {\bf 15}, {\bf 64}, {\bf 175}, $\dots$ for $l=1, 2, 3, \dots$ for both towers of vector modes. The quadratic action with its normalisation constant has been obtained in \cite{Arutyunov:1998hf}. In that paper Arutyunov and Frolov obtained the normalisation by comparison with the covariant action found in reference \cite{DallAgata:1998ahf}. As we shall see, in the holographic dual calculation of the DIS process all the normalisation constants of the supergravity fields are included in a single overall factor. Thus, for the calculation of DIS structure functions that we develop in this work we do not need to obtain the normalisation corresponding to the interaction of two dilatini with the vector field. What really matters is the relative constant $b_{1kj}$ between the minimal coupling interaction and the Pauli interaction terms in the effective five-dimensional action shown in equation (\ref{five-dimensional-action}), that we derive from type IIB supergravity. $b_{1kj}$ is given in terms of certain angular integrals that we solve. In addition, we determine the values of the coupling ${\cal {Q}}$, which as already commented cannot be zero.   Also, for the holographic approach to the DIS we are interested in the interaction with the massless vector modes which are given by setting $l=1$ in equation (\ref{B-modes}). In this case the vector spherical harmonics are Killing vectors of $S^5$. In fact these massless modes are the 15 Yang-Mills gauge fields corresponding to the $SU(4)$ group.   

~

Now, we derive the interaction terms between the vector field $B^l_a$ with $l \geq 1$ and the dilatini. We begin with the covariant EOM of the ten-dimensional dilatino $\hat{\lambda}$, then add second order fluctuations. The EOM reads
\begin{eqnarray}
   \Gamma^M D_M \hat{\lambda}-\frac{i}{2 \ 5!}   \Gamma^{M \dots P} F_{M \dots P} \hat{\lambda} =0 \, . \label{eq-cov-dilatino}
\end{eqnarray}
In a similar way one can obtain the EOM for 
$\hat{\bar{\lambda}}=i \hat{\lambda}^{\dag}\Gamma^{\hat{1}}$. The $\Gamma$ matrices can be written as 
\begin{eqnarray}
    \Gamma^a=\sigma^1\otimes I_4 \otimes \gamma^a, \ \ \ \ \text{and} \ \ \ \ \ \Gamma^{\alpha}=-\sigma^2\otimes \tau^{\alpha} \otimes I_4 \, ,
\end{eqnarray} 
satisfying the Clifford algebra
\begin{eqnarray}
   \{ \Gamma_{\hat{M}},\Gamma_{\hat{N}}\}=2 \eta_{\hat{M} \hat{N}} \, ,  \ \ \ \ \{\gamma_{\hat{a}},\gamma_{\hat{b}} \}=2 \eta_{\hat{a} \hat{b}} \, , \ \ \ \ \{ \tau_{\hat{\alpha}}, \tau_{\hat{\beta}} \}=2 \delta_{\hat{\alpha} \hat{\beta}} \, .
\end{eqnarray} 
$\sigma^1$ and $\sigma^2$ are the Pauli matrices.
Indices $\hat{M},\hat{N},\hat{P},\dots$, $\hat{a},\hat{b},\hat{c},\dots$ and $\hat{\alpha}, \hat{\beta}, \hat{\gamma}, \dots$ correspond to flat space-time in ten and five dimensions, respectively.

The dilatino field is a right-handed spinor 
\begin{equation}
    \hat{\lambda}=\frac{1}{2}\left(1-\Gamma_{11} \right) \hat{\lambda}= \left({\begin{array}{c}
   0\\
   \lambda \\
  \end{array} }\right) \, , \label{dilatino-16}
\end{equation}
where
\begin{equation}
    \Gamma_{11}=\Gamma^{\hat{0}} \dots \Gamma^{\hat{9}}= \left({\begin{array}{cc}
   I_{16} & 0 \\ 
   0      & -I_{16} \\
  \end{array} }\right) \, .
\end{equation}
The covariant derivative $D_M$ is defined in terms of the spin connection and the operator $Q_M$:
\begin{eqnarray}
  D_{M}=\partial_M+\frac{1}{2} \omega_M^{\hat{N} \hat{O}} \Sigma_{\hat{N} \hat{O}}+Q_M \, .
\end{eqnarray}
The $U(1)$ connection $Q_M$ couples the dilatini with the axio-dilaton field, thus it is irrelevant for the DIS process we are interested in at tree level\footnote{It would be relevant for one-loop supergravity calculations of the holographic dual description of DIS which correspond to $\frac{1}{N}$ corrections.}. We define  $\Sigma_{\hat{A}\hat{B}}=\frac{1}{4}(\Gamma_{\hat{A}}\Gamma_{\hat{B}}-\Gamma_{\hat{B}}\Gamma_{\hat{A}})$ while the spin connection is given in terms of the vielbein
\begin{eqnarray}
    \omega_{M}^{\hat{M} \hat{N}}&=&e_{O}^{\hat{M}} \nabla_{M}e^{O \hat{N}}=-e^{O\hat{N}} \nabla_{M} e_{O}^{\hat{M}}  = e_{O}^{\hat{M}} \partial_{M}e^{O \hat{N}}+e_{O}^{\hat{M}}e^{R\hat{N}} \Gamma_{R M}^{O} \, , \label{spin-con-1}
\end{eqnarray}
and
\begin{equation}
    \eta_{\hat{M} \hat{N}} e^{\hat{M}}_{N} e^{\hat{N}}_{M}=g_{M N}.\label{vielbein}
\end{equation}
By taking into account the off-diagonal metric perturbations one obtains the corresponding first-order corrections to the vielbein. We choose the standard parametrization to describe the vielbein and the Kaluza-Klein fields, obtaining 
\begin{eqnarray}
    (e^0)_{N}^{\hat{M}}=\sqrt{g_{NN}} \delta^{\hat{M}}_N  \, , \ \ \ \ \ \ \
    (e^0)_{\hat{M}}^{N}=\frac{1}{\sqrt{g_{NN}}} \delta_{\hat{M}}^N \, .
\end{eqnarray}
The vielbein depends on the metric perturbations in the following way
\begin{eqnarray}
  &&e_{N}^{\hat{N}}=(e^0)_{N}^{\hat{N}}+\delta e_{N}^{\hat{N}}=
  \left({\begin{array}{cc}
   e_a^{\hat{b}} & A^{I_5}_{a} Y^{I_5 \hat{\beta}} \\ 
   0 & e_{\alpha}^{\hat{\beta}} \\
  \end{array} }\right) \, , \ \ \ \ \ \ 
  e_{\hat{N}}^{N}=
  \left({\begin{array}{cc}
   e_{\hat{a}}^{a} & - A^{I_5}_{\hat{a}} Y^{I_5 \beta} \\
   0 & e_{\hat{\alpha}}^{\beta} \\
  \end{array} }\right) \, , \nonumber \\
  &&e^{\hat{N} N}=
  \left({\begin{array}{cc}
   e^{\hat{a} a} & 0 \\
   -A^{I_5 \hat{a}} Y^{I_5 \beta} & e^{\hat{\alpha} \beta} \\
  \end{array} }\right) \, .
\end{eqnarray}
Then, we can collect the first-order terms in the vector fields $A_a$ which contribute to the variation of the vielbein $\delta e$. Firstly, we analyse the fluctuations of the kinetic term in equation (\ref{eq-cov-dilatino}) which come from the contraction with the metric. We have to consider the vielbein present in the $\Gamma^M D_M$ contraction as well as the two vielbeins in the definition of the spin connection and the Christoffel symbol.
We begin with the study of the fluctuations coming from $\Gamma^M D_M$. Thus, we obtain
\begin{eqnarray}
 \Gamma^{\hat{L}}\delta e_{\hat{L}}^M \left(\partial_M+\frac{1}{2} \omega_M^{\hat{M} \hat{N}} \Sigma_{\hat{M} \hat{N}} \right)\hat{\lambda}&=&  -A^{I_5}_{\hat{a}} \Gamma^{\hat{a}} Y^{I_5 \alpha} \left(\partial_{\alpha}+\frac{1}{2}\omega_{\alpha}^{\hat{M} \hat{N}}\Sigma_{\hat{M} \hat{N}} \right)\hat{\lambda} \nonumber \\ &=&-\left(\sigma^1 \otimes I_4 \ Y^{I_5 \alpha}D_{\alpha} \ \otimes \gamma^{a} A^{I_5}_a\right) \hat{\lambda}  \, .
\end{eqnarray}
In the second line of this equation we obtain the factor $\gamma^{a} A^{I_5}_a$ which has indices running on AdS$_5$, which is similar to the minimal coupling. However, there is an additional factor on $S^5$, which is the contraction of the vector spherical harmonic $Y^{I_5 \alpha}$ and the covariant derivative. Recall that when $l=1$, $Y^{I_5 \alpha}$ becomes a Killing vector on $S^5$. Together with another contribution this generates the angular momentum operator associated with the $U(1)$ symmetry. The corresponding spinor spherical harmonics are eigenstates of this operator, and in this way it reduces to the minimal coupling studied in \cite{Polchinski:2002jw}.

Now, let us consider perturbations on the spin connection given by
\begin{eqnarray}
     \Gamma^{\hat{L}} e_{\hat{L}}^M \left(\frac{1}{2} \delta \omega_M^{\hat{M} \hat{N}} \Sigma_{\hat{M} \hat{N}} \right)&=& \frac{1}{2}  \Gamma^{\hat{L}} e_{\hat{L}}^M  \left(\delta e_O^{\hat{M}} \nabla_M e^{O \hat{N}}+ e_O^{\hat{M}} \nabla_M \delta e^{O \hat{N}}
      +e_{N}^{\hat{M}}e^{R\hat{N}} \delta \Gamma_{R M}^{N} \right) \Sigma_{\hat{M} \hat{N}}  \nonumber \\
     &=&
     -\sigma^2\otimes Y^{I_5 \alpha}\tau_{\alpha} \otimes  \left(\frac{1}{4} e^{b\hat{c}}e^{a \hat{b}}\tilde{F}^{I_5}_{ba}\Sigma_{\hat{c}\hat{b}}\right)  \nonumber \\
&&      \ -\sigma^1 \otimes \tau^{\beta}\nabla_{\beta}(Y^{I_5 \alpha})\tau_{\alpha} \otimes \frac{1}{4} \gamma^a A^{I_5}_a \, ,
\end{eqnarray}
where we have used the following identities deduced from the commutation relations of the $\Gamma$ matrices
\begin{eqnarray}
     \Gamma_{\Hat{\gamma}}\Sigma_{\hat{\beta}\Hat{a}}&=&\Sigma_{\Hat{\gamma} \hat{\beta}} \Gamma_{\hat{a}}+\frac{1}{2} \eta_{\Hat{\gamma} \Hat{\beta}}\Gamma_{\Hat{a}} \, ,\\
     \Sigma_{\hat{\beta}\hat{a}}\Gamma_{\hat{\gamma}}&=&\frac{1}{4}(\Gamma_{\hat{\beta}} \Gamma_{\hat{a}}-\Gamma_{\hat{a}} \Gamma_{\hat{\beta}})\Gamma_{\hat{\gamma}}=\Sigma_{\hat{\gamma}\hat{\beta}}\Gamma_{\hat{a}}-\frac{1}{2} \eta_{\hat{\gamma}\hat{\beta}}\Gamma_{\hat{a}} \, .
\end{eqnarray}
As a consistency check one can show that the non-gauge invariant terms of the form $\nabla \cdot A$ which come from the vielbein perturbations cancel exactly with those coming from the Christoffel symbols, thus leading to a gauge invariant result as expected.

Now, let us write all the terms induced by the off-diagonal metric perturbations on $\Gamma^M D_M$ acting on the dilatino. We obtain
\begin{eqnarray}
\Gamma^M D_M \hat{\lambda} &=& ( \gamma^{a} D_a+ i  \tau^\alpha  D_\alpha) \lambda- \gamma^a A^{I_5}_a \left(Y^{I_5 \alpha} D_{\alpha}-\frac{1}{4}\tau^{\alpha} \tau^{\gamma} \nabla_{\gamma}Y^{I_5}_\alpha \right)\lambda \nonumber \\
&& +\frac{i}{4}  \tilde{F}^{I_5 ab} \Sigma_{ab} \tau_{\alpha}
Y^{I_5 \alpha}\lambda +{\cal{O}}(\psi^3) \, , \label{eq-red-cin}
\end{eqnarray}
where we have defined the tensor $\tilde{F}^{I_5}_{ab} = \partial_a A^{I_5}_b - \partial_b A^{I_5}_a$, related to the gauge field coming from the off-diagonal metric perturbation. In  equation (\ref{eq-red-cin}) the term ${\cal{O}}(\psi^3)$ indicates that this expansion includes quadratic terms in the fluctuations of the considered fields. $\psi$ here denotes  linear fluctuations of the type IIB supergravity fields.

Note that in order to obtain the above equation we have expanded the fermionic field $\lambda$ of equation (\ref{dilatino-16}) in terms of the spinor spherical harmonics on $S^5$, $\Theta_k^+(y)$ and  $\Theta_k^-(y)$, 
\begin{eqnarray}
\lambda=\sum_k \left(\lambda^+_k(x) \Theta_k^+(y) + \lambda_k^{-}(x)
\Theta_k^-(y)\right) \, , \label{spinor-expansion}
\end{eqnarray}
which satisfy the corresponding Dirac equations
\begin{eqnarray}
\tau^{\alpha}D_{\alpha} \Theta_k^{\pm}= \mp i
\left(k+\frac{5}{2} \right)\Theta_k^{\pm} \ \ \ \ \ \ \text{with} \ \
\  k \geq 0 \, . \label{thetakpm}
\end{eqnarray}
The construction of the spinor spherical harmonics on $S^5$ can be done iteratively from lower dimensional spheres using the method developed in \cite{Camporesi:1995fb}. Another way to construct them has been studied in reference \cite{vanNieuwenhuizen:2012zk} from Killing spinors on $S^5$.

Also, to obtain equation (\ref{eq-red-cin}) the matrices $\tau_\alpha$ and $\nabla_{\gamma}Y^{I_5}_{\alpha}$ have been exchanged. This only gives a change of sign, since $\eta_{\alpha \tau}$ from the anti-commutation relation contracts the vector spherical harmonic with the covariant derivative and it vanishes. The charge eigenvalues are associated with the operator $v^{\alpha}
D_{\alpha}-\frac{1}{4}\tau^{\alpha} \tau^{\gamma}
\nabla_{\gamma}v_{\alpha} $, where $Y^{\alpha}$ is replaced by the Killing vector $v^{\alpha}$ \footnote{Note that this operator reduces to $i \frac{\partial}{\partial_{\theta_1}}$ for the first constant Killing vector. $\theta_1, \dots , \theta_5$ denote the angles of $S^5$.}.

Now, let us focus on the perturbations of the four-form potential at first order which are relevant for the case we consider. Thus, we have
\begin{eqnarray}
  \frac{i}{2\cdot 5!}  \Gamma^{M1\cdots M_5} F_{M1\cdots M_5}\hat{\lambda} = -(\sigma^+ \otimes
  I_{16}) \hat{\lambda}+ {\cal{O}}(\psi^2) \, , \label{int-5f}
\end{eqnarray}
where $\sigma^+=\frac{\sigma^1+i \sigma^2}{2}$.
At first order it corresponds to a mass term in AdS$_5$ which adds to the D'Alambertian operator coming from the kinetic term. In order to obtain the second order perturbations we note that $F_5$ includes vector perturbations related to the vector components of the four-form potential given in equation (\ref{fluc-5-form}). These perturbations lead to the following contributions to $F_5$: $f^{(1)}_{a \alpha \beta \gamma \delta}$,
$f^{(2)}_{\alpha a b c d}$, $f^{(3)}_{a b \alpha \beta \gamma}$ and $f^{(4)}_{a b c \alpha \beta}$.

We can write the $F_5$ fluctuations by using the decomposition in spherical harmonics which leads to
\begin{eqnarray}
f_{a \alpha \beta \gamma \delta}^{(1)}=5\partial_{[a}  a_{\alpha \beta \gamma \delta]}&=&\partial_a a_{\alpha \beta \gamma \delta} -\partial_{\alpha}a_{a \beta \gamma \delta}-\partial_{\beta}a_{\alpha  a \gamma \delta} - \partial_{\gamma} a_{\alpha \beta a \delta} - \partial_{\delta} a_{\alpha \beta \gamma a} \nonumber \\
&=& \Phi_a^{I_5}(x) \left(-\epsilon_{\beta \gamma \delta \epsilon
\tau}\partial_{\alpha}+\epsilon_{\alpha \gamma \delta \epsilon \tau}
\partial_{\beta}+\epsilon_{\beta \alpha \delta \epsilon \tau}
\partial_{\gamma}+\epsilon_{\beta \gamma \alpha \epsilon \tau}
\partial_{\delta} \right)  \nabla^{\epsilon} Y^{I_5  \tau}(y) .
\nonumber \label{eq-f1} \\
\end{eqnarray}
There is a sum over indices of the spherical harmonics on $S^5$ that we omit to simplify the notation. The rest of the fluctuations can be written as
\begin{eqnarray}
f^{(2)}_{\alpha a b c d}&=&Y_\alpha^{I_5}(y) \left(-\epsilon_{b c d e f}\partial_{a}+\epsilon_{a c d e f} \partial_{b}+\epsilon_{b a d e f} \partial_{c}+\epsilon_{b c a e f} \partial_{d} \right)  \nabla^{e} \Phi^{I_5 f}(x) \, ,\\
f^{(3)}_{ab\alpha \beta \gamma}&=&\left( \partial_a \Phi_b^{I_5}(x)-\partial_b \Phi^{I_5}_a(x) \right) \epsilon_{\alpha \beta\gamma \delta \epsilon} \nabla^{\delta} Y^{I_5 \epsilon}=\bar{F}^{I_5}_{ab} \epsilon_{\alpha \beta\gamma \delta \epsilon} \nabla^{\delta} Y^{I_5 \epsilon}(y) \, , \\
f_{abc\alpha\beta}^{(4)}&=&\partial_{[\alpha}a_{\beta]abc}=\left(\nabla_{\alpha}
Y^{I_5}_{\beta}-\nabla_{\beta}Y^{I_5}_{\alpha}\right) \epsilon_{abcde}\nabla^{d}
\Phi^{I_5 e}(x) \, ,
\end{eqnarray}
where we have defined a new two-form field strength $\bar{F}^{I_5}_{ab}=\partial_a \Phi_b^{I_5}-\partial_b \Phi^{I_5}_a$. After some algebra, and using properties of the gamma matrices we obtain 
\begin{eqnarray}
   \frac{i}{2\cdot 5!} \Gamma^{M_1\cdots M_5} f^{(1)}_{M_1\cdots M_5} &=&  \frac{i \ }{2} \left(\sigma^1 \otimes \left( \tau_{\alpha} \nabla_{\beta} \nabla^{\beta} Y^{I_5 \alpha}-\tau_{\beta}\nabla_{\alpha} \nabla^{\beta} Y^{I_5 \alpha} \right)   \otimes \gamma^a \Phi_a^{I_5} \right) \, , \\
   \frac{i}{2\cdot 5!}  \Gamma^{M_1\cdots M_5} f^{(2)}_{M_1\cdots M_5} &=&  \frac{1}{2} \left(\sigma^2 \otimes  \tau^{\alpha}Y^{I_5}_{\alpha} \otimes \left( \gamma_{a}\nabla_{b} \nabla^{b} \Phi^{I_5 a}-\gamma_{a} \nabla_{b} \nabla^{a} \Phi^{I_5 b} \right) \right)  \, , \\
   \frac{i}{2\cdot 5!}  \Gamma^{M_1\cdots M_5} f^{(3)}_{M_1\cdots M_5} &=&   \frac{i}{2} \left(\sigma^2\otimes \tau^{\alpha}\tau^{\beta}\nabla_{\alpha}Y^{I_5}_{\beta} \otimes \gamma^{a} \gamma^b \bar{F}^{I_5}_{ab} \right) \, ,  \\
   \frac{i}{2\cdot 5!}  \Gamma^{M_1\cdots M_5} f^{(4)}_{M_1\cdots M_5} &=&  \frac{1}{2} \left(\sigma^1\otimes \tau^{\alpha}\tau^{\beta}\nabla_{\alpha}Y^{I_5}_{\beta} \otimes \gamma^{a} \gamma^b \bar{F}^{I_5}_{ab} \right) \, .
\end{eqnarray}

Finally, we consider the contributions of the fluctuations of the vielbein to the contraction of $\Gamma^{\hat{M}\cdots \hat{P}}$ with $F_5$, leading to 
\begin{eqnarray}
i \Gamma^{\hat{M}_5\cdots \hat{M}_5} \delta e^{M_1}_{\hat{M_1}}\cdots e_{\hat{M}_5}^{M_5} F_{M_1\cdots M_5} &+& \dots + i \Gamma^{\hat{M}_5\cdots \hat{M}_5} e^{M_1}_{\hat{M_1}}\cdots \delta  e_{\hat{M}_5}^{M_5} F_{M_1\cdots M_5} \nonumber \\
&=&i \left( \frac{5}{2 \cdot 5!} \Gamma^{\hat{\alpha}}\delta e_{\hat{\alpha}}^a \Gamma^{b} \Gamma^{c}\Gamma^{d} \Gamma^{e} \epsilon_{abcde} +\frac{5}{2 \cdot 5!} \Gamma^{\hat{a}} \delta e_{\hat{a}}^{\beta}\Gamma^{\gamma} \Gamma^{\delta} \Gamma^{\epsilon} \Gamma^{\tau} \epsilon_{\beta \gamma \delta \epsilon \tau} \right)\nonumber \\
 &=&\frac{i}{2 \cdot 4!} \Gamma^{\hat{a}} A_{\hat{a}} Y^{\beta} \Gamma^{\gamma} \Gamma^{\delta} \Gamma^{\epsilon} \Gamma^{\tau} \epsilon_{\beta \gamma \delta \epsilon \tau} \nonumber \\
 &=&-i \sigma^1 \otimes \tau^{\beta}Y^{I_5}_{\beta}\otimes \frac{1}{2} \gamma^{a} A^{I_5}_{a} \, .
\end{eqnarray}

The EOMs of the dilatino field modes $\lambda^\pm_k$ corrected to second order in the perturbations of the fields read 
\begin{eqnarray}
    (\gamma^{a}D_{a} + m^{\pm}_{k} ) \lambda^\pm_{k}&=& \left(   \gamma^a A^{I_5}_a (Y^{I_5 \alpha} D_{\alpha}-\frac{1}{4}\tau^{\alpha} \tau^{\gamma} \nabla_{\gamma} Y^{I_5}_{\alpha} )  -\frac{i}{4}  \tilde{F}^{I_5 ab} \ \Sigma_{ab} \tau_{\alpha} Y^{I_5 \alpha}  \right. \nonumber \\
    &&\left.+
    \frac{i }{2}  \left( \tau_{\alpha} \nabla_{\beta} \nabla^{\beta} Y^{I_5 \alpha}-\tau_{\beta}\nabla_{\alpha} \nabla^{\beta} Y^{I_5 \alpha} \right)   \gamma^a \Phi_a^{I_5} \right. \nonumber \\
&&\   -\frac{i }{2}   \tau^{\alpha}Y_{\alpha}^{I_5}   \left(
\gamma_{a}\nabla_{b} \nabla^{b} \Phi^{I_5 a}-\gamma_{a}
\nabla_{b} \nabla^{a} \Phi^{I_5 b} \right) \nonumber
\\
&& \ \left. +
\left(\tau^{\alpha}\tau^{\beta} \nabla_{\alpha} Y^{I_5}_{\beta}\right)
\gamma^{a} \gamma^b \bar{F}^{I_5}_{ab} -\frac{i}{2}
\tau^{\alpha}Y^{I_5}_{\alpha} \gamma^a A^{I_5}_{a} \right) \lambda^\pm_{k}+
{\cal{O}}(\psi^3) \, ,
\end{eqnarray}
where $m^+_{k}=k+\frac{7}{2}$ and $m^-_{k}=-\left(k+\frac{3}{2}\right)$ are the masses of the Kaluza-Klein dilatino modes $\lambda^\pm_k$. Note that we again have omitted the sum for $\lambda^{\pm}_k \Theta_k^{\pm}$.

\subsection{Interaction terms of two dilatini with a massless vector field mode $B_a^1$}

Now, let us study the perturbations associated with a massless vector mode. Such perturbations correspond to setting $l=1$, therefore the corresponding vector spherical harmonic on $S^5$ is a Killing vector on the sphere:  $Y^{1}_{\alpha}=v_{\alpha}$, which satisfies the equation
\begin{equation}
    \nabla_{\alpha}v_{\beta}+\nabla_{\beta}v_{\alpha}=0 \, .
    \label{killing-vec}
\end{equation}
In order to obtain the interactions with two dilatini we now consider the massless vector mode $B_a^1$. Thus, we obtain 
\begin{eqnarray}
    (\gamma^{a}D_{a}  + m^\pm_{k} ) \lambda^\pm_{k}&=& \left(\frac{1}{3} \gamma^a B^1_a (v^{\alpha} D_{\alpha}-\frac{1}{4}\tau^{\alpha} \tau^{\gamma} \nabla_{\gamma}v_{\alpha} ) -\frac{i}{12}  F^{ab} \ \Sigma_{ab} \tau_{\alpha} v^{\alpha}  \right. \label{eq-int-final-1}  \\
    &&\left.-
    \frac{i \ }{2\cdot 24}  \left( \tau_{\alpha} \nabla_{\beta} \nabla^{\beta} v^{\alpha} - \tau_{\beta}\nabla_{\alpha} \nabla^{\beta} v^{\alpha} \right)   \gamma^a B^1_a \right.  \label{eq-int-final-2}\\
&&\  +\frac{i \ }{2\cdot 24}   \tau^{\alpha} v_{\alpha} \left(
\gamma_{a}\nabla_{b} \nabla^{b} B^{1 a}-\gamma_{a} \nabla_{b}
\nabla^{a} B^{1 b} \right)  \label{interaccion-3}
\\
&& \  \left. - \frac{1}{24}
\left(\tau^{\alpha}\tau^{\beta}\nabla_{\alpha}v_{\beta}\right)
\gamma^{a} \gamma^b F_{ab} -\frac{i}{6}
\tau^{\alpha}v_{\alpha} \gamma^a B^1_{a} \right) \lambda^\pm_{k} \, . \label{last-eq}
\end{eqnarray}
The term in the line (\ref{interaccion-3}) vanishes since it contains the EOM of the massless vector field at first order 
$\gamma_a \nabla_b F^{ba}=0$,
where we have defined a third type of two-form field strength:
\begin{equation}
F_{ab}=\nabla_a B^1_b - \nabla_b B^1_a \, ,
\end{equation}
which turns out to be a linear combination of the form $F_{ab}=\tilde{F}_{ab}^1-16 \bar{F}_{ab}^1$. In fact, the term (\ref{interaccion-3}) introduces cubic corrections which are irrelevant for the calculation we are interested in. The contribution of line (\ref{eq-int-final-2}) can be simplified by noting that the Killing vector $v^{\alpha}$ satisfies the EOM of the vector spherical harmonics for $l=1$, and it exactly cancels the last term in the line (\ref{last-eq}), as explicitly shown below:
\begin{eqnarray}
\frac{-i \ }{2\cdot24}  \left( \tau_{\alpha} \nabla_{\beta}
\nabla^{\beta} v^{\alpha}(y)-\tau_{\beta}\nabla_{\alpha}
\nabla^{\beta} v^{\alpha}(y) \right)  \gamma^a B^1_a(x) &=& \frac{-2i}{2\cdot24}  \left( \tau_{\alpha} \nabla_{\beta} \nabla^{\beta} v^{\alpha}(y) \right)  \gamma^a B^1_a(x) \nonumber \\ 
&=& \frac{-2}{2\cdot24} (-4i \tau_{\alpha} v^{\alpha}) \gamma^a B^1_a(x) \nonumber \\ 
&=&\frac{i }{6} (\tau_{\alpha} v^{\alpha}) \gamma^a B^1_a(x) \, . \label{eq50}
\end{eqnarray}
The operator multiplying to $\gamma^a B^1_a$ in equation
(\ref{eq-int-final-1}) is associated with an angular momentum operator. If the Killing vector $v^{1}$ is associated with the angle $\theta_1$ the operator becomes the usual angular momentum operator $i\frac{\partial}{\partial \theta_1}$. The spinor spherical harmonics are charge eigenstates of this operator and generate the minimal coupling term used in the holographic dual DIS calculations \cite{Polchinski:2002jw}
\begin{eqnarray}
   \left(v^{\alpha} D_{\alpha}-\frac{1}{4}\tau^{\alpha} \tau^{\gamma} \nabla_{\gamma}v_{\alpha} \right) \Theta^\pm_{k} &=&-i{\cal{Q}} \ \Theta^\pm_{k} \, . \label{eigenvalueeqcharge}
\end{eqnarray}

Next, one has to project the spinor on the spinor spherical harmonics $\Theta^{\pm}_k$, expressing the result in terms of angular integrals on $S^5$
\begin{eqnarray}
    (\gamma^{a}D_{a} + m^\pm_{j} ) \lambda^\pm_{j}&=& -i \frac{{\cal{Q}}}{3} a_{1 k j}    \gamma^a B^1_a \lambda^\pm_{k} -\frac{i}{12}b_{1 k j}  F^{ab} \ \Sigma_{ab} \lambda^\pm_{k} \, ,
\end{eqnarray}
where $a_{1 k j}$ and $b_{1 k j}$ correspond to the following angular integrals\footnote{Also, we have to consider the $\pm$ indices related to the fermionic modes $\lambda^\pm_k$.}:
\begin{eqnarray}
a_{1 k j}&=& \int \  d\Omega_5 \ \Theta^{\dag}_{j} \ \Theta_{k} = \delta_{k j} \, , \\
b_{1 k j}&=&\int d \Omega_5 \ \Theta^{\dag}_{j} \ \left(
\tau_{\alpha}v^{\alpha}-i \tau^{\alpha}\tau^{\beta}
\nabla_{\alpha}v_{\beta} \right)\Theta_{k} \, . \label{coeff-b}
\end{eqnarray}
The integral $b_{1 k j}$ can be simplified and rewritten in terms of the first term\footnote{Details are given in
Appendix A.}. In addition, starting from the EOM for $\bar{\hat{\lambda}}$ one arrives to a consistent result. 

In this way the interactions that we obtain do not have higher order derivatives and can be directly derived from an effective five-dimensional action. Thus, it is not necessary to carry out non-linear redefinitions of the fields as discussed in references \cite{Arutyunov:1999en,Skenderis:2006uy}. 

Finally, we obtain the five-dimensional cubic interaction terms,  which are relevant for the holographic dual description of DIS that we are interested in
\begin{eqnarray}
S_{int}&=& K \int dz \ d^4x \sqrt{-g_{AdS_5}} \times \nonumber \\
&& \ \left( i \frac{{\cal{Q}}}{3}
\bar{\lambda}^\pm_{k} \gamma^a B^1_a\lambda^\pm_{k} +i \ \frac{b^{\pm, \pm}_{1 k j}}{12} \bar{\lambda}^\pm_{j} F^{ab}  \Sigma_{ab} \lambda^\pm_k + i  \frac{b^{\mp, \pm}_{1 k j}}{12} \bar{\lambda}^\pm_{j} F^{ab} \Sigma_{ab} \lambda^\mp_k\right) \, . \label{five-dimensional-action} 
\end{eqnarray}
The first term is similar to the minimal coupling given in \cite{Polchinski:2002jw}, though now we explicitly indicate that the massless vector field is a linear combination of the AdS$_5$ vector components of metric and the four-form potential. Also, this term displays the interactions for all possible dilatino modes labeled with the subindex $k$. Recall that ${\cal{Q}}$ depends on each spinor and cannot be zero. This is a crucial difference with respect to \cite{Gao:2010qk}, and it comes from the fact that equation (\ref{five-dimensional-action}) is directly derived from type IIB supergravity. On the other hand, for the Pauli term an interesting comment is that $b^{\pm,\pm}_{1kj}$ and $b^{\mp, \pm}_{1kj}$ are constants obtained from angular integrals on $S^5$, which we calculate in the following sections and in Appendix A for the case of the twist-3 operator. Another important remark is that the Pauli interaction allows for the mixing of dilatino modes of the two towers ({\it i.e.} $\lambda^{\pm}$) with labels $k$ and $k\pm 1$ through the interaction given by the third term as explained in Section 4.3.2. Also notice that since this effective action contains both kind of dilatino modes $\lambda^{\pm}$, it permits to investigate the contributions of both ${\cal {O}}^{(6)}_k$ and ${\cal {O}}^{(13)}_k$ ${\cal {N}}=4$ SYM theory operators to the current-current OPE which leads to the hadronic tensor.

Although the normalisation constant $K$ can be obtained, for the DIS process we consider in the present work it is not important since there will be other normalisation constants related to the wave-functions of the dilatino and the massless vector field, which we can write as a single overall factor for the expression of the hadronic tensor, thus being the same for all the corresponding structure functions. Therefore, effectively there will be just one single parameter for the whole set of structure functions, namely $|a_0|$. 
%On the other hand, we keep the $N$ dependence explicit since if %we consider the canonical normalisation of the supergravity %fields, the cubic terms have  
%the $1/N^2$ dependence on $N$. 
%We would like to emphasize that we keep all the fields in the %dimensional reduction, and we do not want to decouple some %contributions in order to make a consistent truncation. 
The holographic dual description of the DIS process that we consider requires tree-level Feynman-Witten diagrams involving spin-1/2 hadrons and the electromagnetic current.

\section{Hadronic tensor of spin-1/2 hadrons}

\begin{figure}[h]
\centering
\includegraphics[scale=0.4]{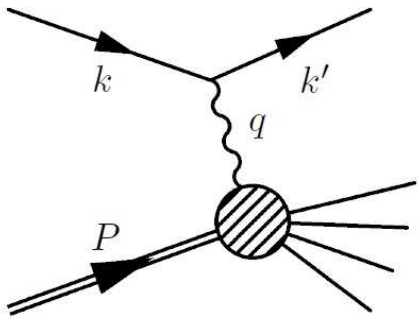}
\caption{{\small Illustration of DIS. A charged lepton exchanges a virtual photon with a hadron of four-momentum $P$. The incoming and outgoing lepton four-momenta are $k$ and $k'$, respectively.}} 
\label{fig-DIS0}
\end{figure}

The structure of hadrons can be characterized by the hadronic tensor $W_{\mu \nu}$ associated to deep inelastic scattering. In this process a charged lepton interacts with a hadron by the exchange of a virtual photon with four-momentum $q^\mu$, the photon probes the hadron which decays into the final states ${\cal {X}}$. Schematically this process is represented in figure \ref{fig-DIS0}, where $P^\mu$ denotes the four-momentum of the initial hadron and $P^\mu_{\cal {X}}$ denotes the total four-momentum of the final hadron states.

The relevant kinematic variable is the Bjorken parameter $x=-q^2/(2 P \cdot q)$ which is kept fixed in the DIS limit $q\to \infty$. The hadronic tensor can be written in terms of the sum of symmetric and antisymmetric parts
\begin{equation}
    W_{\mu \nu}= W^{(S)}_{\mu \nu}(q,P)+i \ W^{(A)}_{\mu \nu}(q,P,s) \, ,
\end{equation}
and for an incident hadron with spin $1/2$ the hadronic tensor can be decomposed in eight structure functions \cite{Anselmino:1994gn,Lampe:1998eu}
\begin{eqnarray}
 W^{(S)}_{\mu \nu}&=&  \left(\eta_{\mu\nu} - \frac{q_\mu q_\nu}{q^2}\right)\left[F_1(x,q^2)+\frac{M \ S \cdot q}{2 P \cdot q} g_5(x,q^2) \right] \nonumber\\
&& - \frac{1}{P\cdot q} \left(P_\mu - \frac{P \cdot q}{q^2} q_\mu \right)\left(P_\nu - \frac{P \cdot q}{q^2} q_\nu \right) \left[F_2(x,q^2) +\frac{M \ S \cdot q}{P\cdot q} g_4(x,q^2) \right] \nonumber\\
&&-\frac{M}{2 P\cdot q} \left[ \left(P_\mu - \frac{P \cdot q}{q^2} q_\mu \right)\left(S_{\nu}-\frac{S \cdot q}{P\cdot q} P_{\nu} \right)+\left( P_\nu - \frac{P \cdot q}{q^2} q_\nu \right)\left(S_{\mu}-\frac{S \cdot q}{P\cdot q} P_{\mu} \right)\right] \nonumber\\
&& g_3(x,q^2) \, ,
\\
 W^{(A)}_{\mu \nu}&=&- \frac{M \ \epsilon_{\mu \nu \rho \sigma } \ q^{\rho}}{P\cdot q} \left(S^{\sigma}g_1(x,q^2) +\left[ S^{\sigma}- \frac{S\cdot q}{P \cdot q} P^{\sigma} \right]g_2(x,q^2)  \right)- \frac{\epsilon_{\mu \nu \rho \sigma} q^{\rho} P^{\sigma}}{2 P \cdot q} F_3(x,q^2) \, , \nonumber \\
\end{eqnarray}
where $M$ is the mass of the incident hadron and $S^\mu$ its spin vector defined as
\begin{equation}
    2S^{\mu}= \bar{u}(k,s) \gamma^{\mu} \gamma_5 u(k,s) \ .
\end{equation}
We are using the mostly plus metric $\eta_{\mu\nu}=\text{diag}(-1,1,1,1)$. It is worth mentioning that in QCD the structure functions  $g_3$, $g_4$, $g_5$ and $F_3$ vanish for the electromagnetic DIS. However, in this work we consider an IR deformation of the chiral theory ${\cal{N}}= 4$ SYM, therefore it leads to a non-vanishing $F_3$ and $g_i$ with $i=3, 4, 5$.

Using the optical theorem one obtains the relations
\begin{eqnarray}
 W^{(S)}_{\mu \nu}=2 \pi \  \text{Im}(T^{(S)}_{\mu\nu})\, , \ \ \ \ \ \ \ \  W^{(A)}_{\mu \nu}=2 \pi \ \text{Im}(T^{(A)}_{\mu\nu})\, ,
\end{eqnarray}
where the tensor $T^{\mu\nu}$ is defined by the time-ordered expectation value of two electromagnetic currents inside the hadron
\begin{equation}
 T^{\mu \nu} \equiv  i \int d^4 \xi \ e^{i q \cdot \xi} \  \langle P ,{\cal{Q}}, S|{\hat{T}} \{J^{\mu}(\xi) J^{\nu}(0)\}|P, {\cal{Q}}, S\rangle \, ,
\end{equation}
while its imaginary part can be expressed as
\begin{eqnarray}
\text{Im}(T^{\mu\nu})&=&2 \pi^2 \sum_{{\cal {X}}} \delta(M_{\cal {X}}^2+(P+q)^2) \langle P ,{\cal{Q}},S| J^{\nu}(0)  | P_{\cal {X}}, {\cal{Q}},S\rangle\langle P_{\cal {X}} ,{\cal{Q}},S| J^{\mu}(0)  | P, {\cal{Q}},S\rangle \label{optical-teo} \, , \nonumber \\
\end{eqnarray}
where $J_{\mu}$ is the electromagnetic current inside the hadron and the tensor can be expressed in terms of a sum over the intermediate states ${\cal {X}}$. The optical theorem has been used to obtain equation (\ref{optical-teo}).

\section{Structure functions of polarised spin-1/2 hadrons from type IIB supergravity}

In this section we carry out the holographic dual calculation of all the structure functions corresponding to the DIS process of a charged lepton off a polarised spin-1/2 hadron which is represented by a dilatino field mode. Since we consider the regime $\lambda_{SYM}^{-1/2} \ll x < 1$ of the Bjorken parameter we use type IIB supergravity compactified on AdS$_5 \times S^5$. The starting point is the five-dimensional action that we have derived in Section 2 given by equation (\ref{five-dimensional-action}). Essentially, it contains two very different types of terms, namely: the first one corresponding to the minimal coupling, and the second one which is the so-called Pauli term (the second and third terms of (\ref{five-dimensional-action})). There are several crucial differences with respect to the calculations of references  \cite{Gao:2009ze} and \cite{Gao:2010qk}. The first one is that we have derived all fermionic interactions from first principles using type IIB supergravity as explained in Section 2. This allows us to calculate the relative constants $b_{1kj}$'s between the terms of  equation (\ref{five-dimensional-action}). Therefore, we obtain the precise contribution of these terms to the hadronic tensor and the structure functions. Also, from the angular integrals on $S^5$ we obtain selection rules (previously unknown) that in the case of the minimal coupling preserve the nature of the incident hadron. Interestingly, in the case of the Pauli interaction the selection rules give also a mixing of the initial and the intermediate states with certain specific quantum numbers, in addition to the case where the initial and intermediate states are the same.

The gauge/string duality relates the large $N$ limit of the conformal $SU(N)$ ${\cal{N}}=4$ SYM theory with type IIB superstring theory on AdS$_5 \times S^5$. In order to study the DIS procces we need to break the conformal invariance by introducing a confinement scale $\Lambda$. Thus, it leads to the so-called hard-wall model proposed by Polchinski and Strasler \cite{Polchinski:2002jw}, where the conformal symmetry is broken by introducing a cut-off $z_0=1/\Lambda$ in the fifth coordinate of AdS$_5$. This leads to a mass gap for the hadrons.

The electromagnetic current inside the hadron comes from gauging the global $U(1)_R$ symmetry subgroup of the $SU(4)_R$ R-symmetry group of ${\cal{N}}=4$ SYM theory. The electromagnetic current operator inserted at the boundary of the AdS$_5$ induces a non-normalizable $B^1_a(x)$ gauge field Kaluza-Klein mode which propagates in the bulk. As explained in Section 2 the vector field $B_a^1(x)$ is associated with a fluctuation of the metric and the four-form potential of the type IIB supergravity. The $U(1)_R$ is dual to an $U(1)$ subgroup of the $SO(6)$ isometry group of the $S^5$, with Killing vector $v_\alpha$. Recall that $x_a$ with $a=0, 1, \dots, 4$ are coordinates on AdS$_5$, then we split $(x_a)=(x_\mu, z)$ where Greek indices $\mu, \nu, \dots =0, 1, 2, 3$ and $z$ is the fifth coordinate. 

In the Lorentz-like gauge
\begin{eqnarray}
\partial^{\mu}B^1_{\mu}+z \partial_z\left( \frac{B^1_z}{z} \right)=0 \, ,
\end{eqnarray}
the vector field satisfies the following Maxwell-Einstein equations in AdS$_5$
\begin{eqnarray}
-q^2 B^1_{\mu} + z \partial_z \left(\frac{1}{z} \partial_z B^1_{\mu} \right)&=&0 \, ,\\
-q^2 B^1_z +\partial_z\left( z \partial_z\left( \frac{B^1_{z}}{z}\right) \right) &=&0 \, ,
\end{eqnarray}
with the boundary condition
\begin{equation}
    B^1_{\mu}(x_\nu,z\to0)=B^1_{\mu}(x_\nu)|_{4d}= n_{\mu}e^{i q \cdot x}
\end{equation}
where $q \cdot x=q_\nu x^\nu$. The solutions of the above equations with the boundary condition are given by
\begin{eqnarray}
B^1_{\mu}=n_{\mu} e^{i q \cdot x} q z K_1(q z), \ \ \ \ \ \  B^1_z=i \  n\cdot q \ e^{i q \cdot x} z K_{0}(q z) \, .
\end{eqnarray}

The Pauli term contains the $F_{\mu \nu}$ tensor, which can be simplified by using the recurrence properties of the Bessel functions, obtaining the following expressions in terms of the modified Bessel functions of second kind, 
\begin{eqnarray}
F_{\mu \nu}&=&e^{i q \cdot x} \ q \ z \ i \, \left(q_{\mu}n_{\nu}-q_{\nu} n_{\mu}\right)K_1(qz) \, ,\\
F_{z \mu}&=& e^{i q \cdot x} \ \left( -n_{\mu} q^2+(n\cdot q) q_{\mu} \right) z K_0(qz) \, .
\end{eqnarray}
The holographic spin-1/2 hadron corresponds to the dilatino field in type IIB supergravity. From equation (\ref{spinor-expansion}) we can expand the spinors in terms of spherical harmonics $\Theta^{\pm}_k(y_{\alpha})$ which are solutions of the Dirac operator with positive or negative eigenvalues (\ref{thetakpm}). These generate two Kaluza-Klein mass towers for spin-1/2 fermions $\lambda^{\pm}_k$ in AdS$_5$ as described in Section 2.1. The $\lambda^-_k(x_{\mu},z)$ modes are the holographic dual fields corresponding to the ${\cal{O}}^{(6)}_k \sim \text{tr}(F_+ \lambda_{{\cal {N}}=4} X^k)$ operators with conformal dimensions $\Delta=k+\frac{7}{2},$ (twist $\tau = k+3$) with $k=0,1,2\cdots$, which belong to the ${\bf 4^*}$, ${\bf 20^*}$, ${\bf 60^*}$, $\cdots$, representations of the $SU(4)_R$. The $\lambda^+_k(x_{\mu},z)$ modes are associated with the ${\cal{O}}^{(13)}_k \sim \text{tr}(F_+^2 \bar{\lambda}_{{\cal {N}}=4} X^k)$ operators of ${\cal{N}}=4$ SYM theory with $\Delta=k+\frac{11}{2}$ (twist $\tau =k+5$), which belong to the ${\bf 4}$, ${\bf 20}$, ${\bf 60}$, $\cdots$, representations of the R-symmetry group. Properties of these operators are summarized in the table 7 of reference \cite{DHoker:2002nbb}\footnote{Notice that in the table of page 50 of \cite{DHoker:2002nbb} the conjugate irreducible representations of $SU(4)_R$ are related to the operators ${\cal {O}}_k^{(13)}$. However, those representations correspond to ${\cal {O}}_k^{(6)}$ operators, and reciprocally (see \cite{Kim:1985ez}).}. 

We consider the $\lambda^-(x_{\mu},z)$ mode with $k=0$ since it has the minimal twist $\tau=3$, therefore providing the leading contribution to the hadronic tensor of spin-1/2 fermions. The $\lambda^-_k(x_\mu, z)$ dilatino mode in AdS$_5$ satisfies the Dirac equation\footnote{Note that in reference \cite{Mueck:1998iz} the authors consider the Euclidean case without a cut-off, therefore their solutions are different.} with mass $\tilde{m}_1(k)=k+\frac{3}{2}$, 
\begin{eqnarray}
\left(\gamma^{m}D_{m}-\tilde{m}_{1}(k)\right)\lambda_k^-= \left(z \gamma^{m} \partial_{m}-2 \gamma^5-\tilde{m}_1(k)\right)\lambda^-_k=0 \, . \label{dirac-dif}
\end{eqnarray}
The normalisable solution with four-momentum $P^{\mu}$ is
\begin{equation}
        \lambda^-_k(x_\mu,z) = C e^{i P \cdot x}z^{\frac{5}{2}}\left(  J_{\tilde{m}_1-\frac{1}{2}}(M z) a_{+}+  J_{\tilde{m}_1+\frac{1}{2}}(M z)a_{-} \right) \, ,
\end{equation} 
where $a_+$ and $a_-$ are spinors satisfying $\gamma_5 a_{\pm}=\pm a_{\pm}$ and $C$ is a normalisation constant. This solution solves the Bessel differential equation obtained after acting with $\gamma^n \partial_n$ on the Dirac equation (\ref{dirac-dif}). Then, using the Dirac equation we find that these spinors are related by $a_{+}=i \frac{\gamma^\mu p_\mu}{M} a_{-}$ . We construct the Dirac spinor in four dimensions $u_{\sigma}$ from $a_\pm = P_{\pm}u_{\sigma}$, therefore the solution can be written as
\begin{equation}
\lambda_k^-(x_\mu,z) = C e^{i P \cdot x}z^{\frac{5}{2}}\left(  J_{\tau-2}(M z) P_{+} +  J_{\tau-1}(M z)P_{-} \right)u_{\sigma} \, , \label{spinor-pos}
\end{equation} 
where 
\begin{equation}
   \gamma^{\mu} P_{\mu} u_{\sigma}= i M u_{\sigma} \, ,  \ \ \ \ \ \ \ P^2=-M^2 \, , \ \ \ \ \ \text{the projectors} \ \ \ P_{\pm}=\frac{(I\pm\gamma^5)}{2}  \, ,
\end{equation}
and $\tau= \Delta-\frac{1}{2}=\tilde{m}_1 + 3/2$ is the twist of the corresponding ${\cal {N}}=4$ SYM theory operator ${\cal {O}}_k^{(6)}$. $I$ denotes the identity matrix.

The modified Bessel function of second kind of $B^1 _{\mu}$ falls exponentially for $\frac{1}{q}<z$. Therefore, in the limit of hard scattering ($\Lambda \ll q$) the interaction occurs in the conformal region near the boundary,
\begin{equation}
    z_{int}\sim \frac{1}{q} \, .
\end{equation}
In this region, we can expand the wave-function of the incident hadron in the DIS limit ($\Lambda \sim M\ll q$) up to second order in $M/q$, obtaining
\begin{eqnarray}
\lambda^-_i \sim e^{i P_i \cdot x} c_i' z_0^{3/2}\left(\frac{z}{z_0}\right)^{\tilde{m}_i+2}\left[P_+ + \frac{M_i z}{ 2(\tilde{m}_1+\frac{1}{2})}P_- \right]u_{\sigma i} \, . \label{lambda0menos}
\end{eqnarray}
In order to obtain the polarised structure functions we have expanded up to second order the initial hadron wave-function. In the non-polarised case of reference \cite{Polchinski:2002jw} this is not necessary. For the intermediate state, the approximation does not hold since $M_{\cal {X}}\sim q$ and we have to use the complete wave function, 
\begin{equation}
    \bar{\lambda}^-_{\cal {X}}=e^{-i P_{\cal {X}} \cdot x}c_{\cal {X}}' \frac{M_{\cal {X}}^{1/2}}{ z_0^{1/2}} z^{5/2} \bar{u}_{\sigma {\cal {X}}}\left[ P_- J_{\tilde{m}_{\cal {X}} -1/2}(M_{\cal {X}} z)+P_+ J_{\tilde{m}_{\cal {X}}+1/2}(M z)    \right] \, . \label{lambdamenos}
\end{equation}
The selection rules derived from the angular integrals allow the interaction with the fermionic states of the $\textbf{4}$, $\textbf{20}$, $\textbf{60}$, $\dots$ representations of $SU(4)$ (recall that in terms of the SYM theory these correspond to ${\cal{O}}^{(13)}_{k}$ operators). The Dirac equations in AdS$_5$ for these supergravity fields are
\begin{equation}
    \left(\gamma^{m}D_{m}+\tilde{m}_2(k)\right)\lambda^+_k= \left(z \gamma^{m} \partial_{m}-2 \gamma^5+\tilde{m}_2(k)\right)\lambda^+_k=0 \, ,
\end{equation}
where $\tilde{m}_2(k)=k+\frac{7}{2}$. Analogously, we calculate the wave-function of the intermediate state with four-momentum $P_{\cal {X}}^{\mu}$, given by
\begin{equation}
\bar{\lambda}^+_{\cal {X}}(x_\mu,z) =   e^{-i P_{\cal {X}} \cdot x} c_{\cal {X}}' \frac{M_{\cal {X}}^{1/2}}{z_0^{1/2}}  z^{\frac{5}{2}}\bar{u}_{\sigma {\cal {X}}} \left(  J_{\tilde{m}_{\cal {X}}+\frac{1}{2}}(M_{\cal {X}} z) P_-+  J_{\tilde{m}_{\cal {X}}-\frac{1}{2}}(M_{\cal {X}} z)P_{+} \right) \, , \label{lambdamas}
\end{equation} 
where the spinor $\bar{u}_{\sigma {\cal {X}}}$ satisfies the Dirac equation.

\subsection{Selection rules for an incident $\tau=3$ spin-1/2 fermion}

The interaction vertices we consider have coefficients involving integrals over the spinor spherical harmonics. These integrals lead to selection rules for the outgoing fermionic states (or the intermediate states in the related forward Compton scattering) and the type of interactions that occur in the AdS$_5$ space. In the strongly coupled ${\cal {N}}=4$ SYM theory the leading contribution to the hadronic tensor comes from the twist $\tau=3$ operator, therefore $k=l_5=0$\footnote{Notice that $k \geq 0$ labels the $\lambda^{\pm}_k$ dilatino modes, while $l\geq 1$ in section 2.1 denotes the $B_a^l$ modes which in the massless case corresponds to $l=1$ (see also footnote 9).} and the other quantum number $l_1, l_2, l_3$ and $l_4$ vanish too, since they satisfy $l_5 \geq l_4 \geq l_3 \geq l_2 \geq l_1$.

Following the formalism proposed in reference \cite{Camporesi:1995fb} we obtain the explicit expressions of the spinor spherical harmonics. The case with minimal twist\footnote{Details of the construction of spinor spherical harmonics and higher twist operators will be reported in \cite{Newpaper}.} has a degeneration related to the $\textbf{4*}$ representation and it implies that there are four spin-1/2 fermionic modes $\lambda^-_{0_a}$ with $a=1, 2, 3, 4$. We have explicitly verified that the final result does not depend on the choice of the initial state (among the above $\lambda^-_{0_a}$ modes belonging to the $\textbf{4*}$ irrep). Thus, without loss of generality we choose the following normalised state
\begin{equation}
    \Theta^{-}_{(0,0,0,0,0)_{a=1}}= 
\frac{e^{-i{\cal{Q}} \theta_1}}{\pi^{3/2}} 
\begin{bmatrix} e^{-i \frac{1}{2}( \theta_3 -\theta_5)} 
\cos (\frac{\theta_2}{2}) \cos (\frac{\theta_4}{2}) \\ 
-e^{i\frac{1}{2}( \theta_3 +\theta_5)}\sin (\frac{\theta_2}{2}) 
\cos (\frac{\theta_4}{2}) \\ -e^{-i\frac{1}{2}( \theta_3 
+\theta_5)}\cos (\frac{\theta_2}{2}) \sin (\frac{\theta_4}{2}) 
\\e^{-i\frac{1}{2}( -\theta_3 +\theta_5)} \sin (\frac{\theta_2}{2}) 
\sin (\frac{\theta_4}{2}) \end{bmatrix} \, ,
\end{equation}
where the charge is ${\cal{Q}}=\frac{1}{2}$ and the sub-index $a$ is associated with the ${\bf 4^*}$ representation of $SU(4)$. In the present case we have chosen $a=1$. The angles $\theta_i$'s correspond to $S^5$ and they are associated with the $l_i$'s.

Now, we carry out the integration between states within the same representation {\bf 4$^*$}. This is related to the Pauli term connecting $\lambda^-_{k=0}$ with $\lambda^-_{k=0}$. 
The only case with a non-vanishing integral corresponds to the coupling with a fermionic state with the same twist 3, which leads to
\begin{equation}
    \int d \Omega_5  (\Theta^{-}_{(0,0,0,0,0)_{a=1}})^{\dag} \tau_{\alpha}v^{\alpha}\Theta^-_{(0,0,0,0,0)_{a=1}} =-\frac{1}{3} \, .
\end{equation}
This coupling is very important because when we calculate the associated matrix element, this contribution must be added to the standard minimal coupling contribution, and as we shall see, this has a very important effect on the structure functions. 
Then, in the calculation of the hadronic tensor (equation (\ref{optical-teo})) the matrix element is multiplied by its conjugate and for this reason, in addition to the contributions found in the reference \cite{Gao:2010qk}, there will be a mixed contribution, involving a minimal coupling vertex on one side and a Pauli interaction vertex on the other side of the forward Compton scattering diagram, which in fact modifies the structure functions. In Section 5 we discuss explicitly the effects of each term for $\tau=3$.

On the other hand, the coupling between $\lambda^-$ states in the \textbf{4$^*$} representation with $\lambda^+$ states in the \textbf{4} representation is controlled by the following integrals which come from the third term in the effective action (\ref{five-dimensional-action})
\begin{eqnarray}
\int d \Omega_5  (\Theta^{+}_{(1,0,0,0,0)_{a=1}})^{\dag} \tau_{\alpha}v^{\alpha}\Theta^-_{(0,0,0,0,0)_{a=1}} &=& -\frac{1}{3 \sqrt{5}} \, ,  \label{integral-1} \\ 
\int d \Omega_5  (\Theta^{+}_{(1,1,0,0,0)_{a=1}})^{\dag} \tau_{\alpha}v^{\alpha}\Theta^-_{(0,0,0,0,0)_{a=1}} &=& \frac{1}{\sqrt{30}} \, , \label{integral-2} \\
\int d \Omega_5  (\Theta^{+}_{(1,1,1,0,0)_{a=3}})^{\dag} \tau_{\alpha}v^{\alpha}\Theta^-_{(0,0,0,0,0)_{a=1}} &=& \frac{1}{3 \sqrt{2}}, \, \label{integral-3}  \\  
\int d \Omega_5  (\Theta^{+}_{(1,1,1,1,0)_{a=3}})^{\dag} \tau_{\alpha}v^{\alpha}\Theta^-_{(0,0,0,0,0)_{a=1}} &=& -\frac{1}{3}\, , \label{integral-4}
\end{eqnarray}
where the sub-index $a=3$ represents a state associated with the ${\cal {O}}_{k=1}^{(13)}$ operator. This corresponds to $l_1=0$. In this case the corresponding fermionic states have both the same ${\cal {Q}}$, namely: ${\cal {Q}}_1={\cal {Q}}_3=\frac{1}{2}$. These are the only non-vanishing integrals.

\begin{figure}[h]
%\centering
\includegraphics[scale=1]{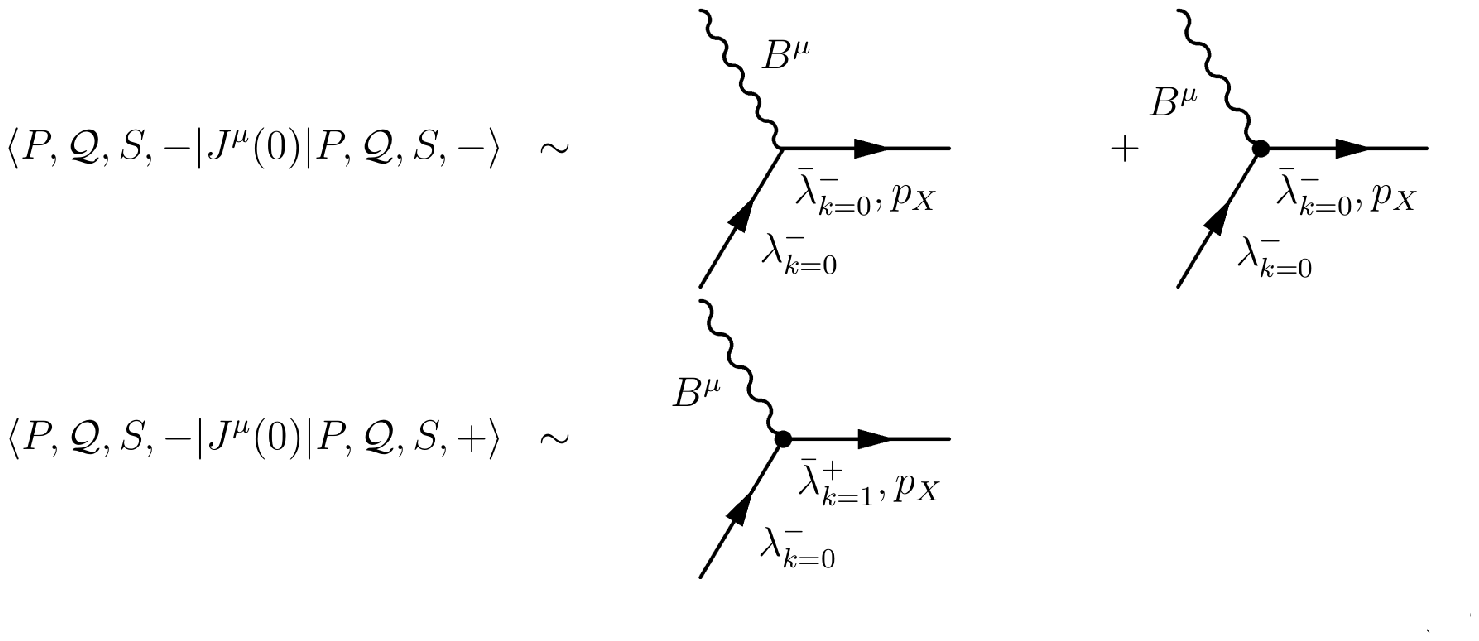}
\caption{{\small On the left we write the matrix elements of the electromagnetic current inside the hadron. On the right we draw the corresponding holographic dual Feynman diagrams. The diagram in the second line shows an example of the $\lambda_k^-$ and $\lambda_{k+1}^+$ mixing.}} 
\label{fig-sk-DIS}
\end{figure}

In figure \ref{fig-sk-DIS} we detail the Feynman diagrams corresponding to the matrix elements of the electromagnetic current inside the hadron in terms of their dual type IIB supergravity fields representation. These diagrams are the building blocks to construct the corresponding forward Compton scattering Feynman diagrams which allow to derive the hadronic tensor.
The first diagram, which connects the initial state with same final state, corresponds to the minimal coupling. The other diagrams with dotted vertices correspond to the Pauli interactions. The $-(+)$ sign in the kets denote the representation $\textbf{4*}$ ($\textbf{4}$) of $SU(4)_R$ of the dual SYM operators, respectively.

\subsection{The minimal coupling contributions}

Let us consider an incident hadron of (four-dimensional) mass $M_i$, whose holographic dual representation is given by a Kaluza-Klein mode of mass\footnote{Recall that $m_i$ is the mass of the Kaluza-Klein mode coming from the dimensional reduction of the dilatino field on $S^5$. Here we consider $\tilde{m}_i \equiv -m_i$ which is positive for the ${\cal {O}}_k^{(6)}$ operator with twist $k+3$, since $m_i=-k-3/2$ with $k \geq 0$ in this case.} $m_i$ from the spontaneous compactification of type IIB supergravity on AdS$_5 \times S^5$.
Its wave-function has been obtained in equation (\ref{spinor-pos}). We consider the limit $M_{hadron}^2 \ll q^2$, which is consistent within the DIS context\footnote{$M_{hadron}$ represents the mass of the incident $M_i$, intermediate $M_{\cal {X}}$ and final hadron $M_f$, in the forward Compton scattering.}, and the solution is approximated at order $M_{hadron}^2/q^2$. We need to keep an additional order in comparison with the non-polarised case, thus the wave-function is given in equation (\ref{lambda0menos}).
The spinor $\lambda^-_i$ is coupled to another spinor 
$\bar{\lambda}^-_{\cal {X}}$ of Kaluza-Klein mass $m_{\cal {X}}$ representing an intermediate hadron of four-dimensional mass $M^2_{\cal {X}}=-(P+q)^2$. Now, since the mass $M_{\cal {X}}$ is of the same order as $q$, we cannot use the same approximation as for the incident hadron. Thus, we must use the complete solution (\ref{lambdamenos}). Next, we consider the contribution to the five-dimensional action (\ref{five-dimensional-action}) from the minimal coupling split in two integrals
\begin{eqnarray}
I_1+I_2=i \int dz \ d^4x  \ \bar{\lambda}^-_{k} \gamma^a B^1_a \lambda^-_{k}  \, , \label{five-dimensional-action-1} 
\end{eqnarray}
where $I_1$ and $I_2$ correspond to the integrals of the $\bar{\lambda}^-_{k} \gamma^\mu B^1_\mu \lambda^-_{k}$ and $\bar{\lambda}^-_{k} \gamma^z B^1_z \lambda^-_{k}$ components, respectively. The explicit calculations are given in Appendix B.

Then, the matrix elements of $\langle J^{\mu} \rangle$ are given by the following expressions
\begin{eqnarray}
&& n_{\mu}\langle J^{\mu} \rangle = \frac{i}{2\pi}\left(\frac{M_{\cal {X}}}{z_0}\right)^{\frac{1}{2}} \times \nonumber \\
&& \left(c_1q \bar{u}_{\sigma {\cal {X}}}\slashed{n} P_+ u_{\sigma i} + c_2 q \bar{u}_{\sigma {\cal {X}}}\slashed{n} P_- u_{\sigma i}-c_3 \left( i n \cdot q \right)  \bar{u}_{\sigma {\cal {X}}} P_-u_{\sigma i} + c_4  \left( i n \cdot q \right) \bar{u}_{\sigma {\cal {X}}}P_+u_{\sigma i}    \right) , \\
&& n_{\nu} \langle J^{\nu *} \rangle =- \frac{i}{2\pi}\left(\frac{M_{\cal {X}}}{z_0}\right)^{\frac{1}{2}} \times \nonumber \\
&& \left(c_1 q  \bar{u}_{\sigma i} \slashed{n} P_+ u_{\sigma {\cal {X}}} + c_2 q \bar{u}_{\sigma i} \slashed{n} P_- u_{\sigma {\cal {X}}} - c_3 \left( i n \cdot q \right)  \bar{u}_{\sigma i} P_+ u_{\sigma {\cal {X}}}+ c_4  \left( i n \cdot q \right) \bar{u}_{\sigma i}P_-u_{\sigma {\cal {X}}}    \right) , 
\end{eqnarray}
where the constants $c_1, c_2, c_3$ and $c_4$ are obtained from the integrals of the Bessel functions, and they are defined in Appendix B. For the minimal coupling derived from the spontaneous compactification of type IIB supergravity on AdS$_5 \times S^5$ we find the following selection rule in terms of the twist of the corresponding ${\cal {N}}=4$ SYM theory operators: $\tau_i=\tau_{\cal {X}} \equiv \tau$. This rule straightforwardly derives from the corresponding angular integral on $S^5$. It means that there is no mixing between the initial and the intermediate hadrons, thus preserving the nature of the incident hadron through the tree-level diagram, either thinking of DIS or forward Compton scattering processes.

Then, by summing over the intermediate states we obtain the  hadronic tensor  
\begin{eqnarray}
 n_{\mu} n_{\nu}   W^{\mu \nu}&=&n_{\mu} n_{\nu}  \ 2 \pi^2 \ \sum_{{\cal {X}}}\delta(M_{\cal {X}}^2+(P+q)^2) \langle J^{\mu}(0) \rangle \langle J^{*\nu}(0) \rangle \nonumber \\
 &\sim& \frac{z_0}{2\pi M_{\cal {X}}} n_{\mu} n_{\nu} \langle J^{\mu}(0) \rangle \langle J^{*\nu}(0)\rangle \, ,
\end{eqnarray}
and from it we derive the contribution to all the structure functions corresponding to the minimal coupling exclusively. We use the superscript $m$ to denote minimal coupling,
\begin{eqnarray}
F^m_1&=&\frac{F_2^m}{2}=\frac{F_3^m}{2}=g_1^m=\frac{g_3^m}{2}=\frac{g_4^m}{2}=\frac{g_5^m}{2}=c_1^2 q^2 p \cdot q \nonumber \\
&=& \frac{|a_0|^2}{8} \ \Gamma^2(\tau)  \left(\frac{\Lambda^2}{q^2}\right)^{\tau-1} x^{\tau+1}(1-x)^{\tau-2}  \, , \\
g_2^m&=& \frac{q^5}{4 x^2} \left(c_2 c_4-c_1 c_3  \right) =\left( \frac{1}{2}\frac{\tau+1}{\tau-1}-\frac{x \tau}{\tau-1} \right) \frac{|a_0|^2}{8} \ \Gamma^2(\tau) \left(\frac{\Lambda^2}{q^2}\right)^{\tau-1} x^{\tau}(1-x)^{\tau-2} \, ,
\end{eqnarray}
having defined the constant $a_0=2 \pi c'_i c'_{\cal {X}} 2^\tau K$. Since we have factorized out $\beta_m=\frac{{\cal {Q}}}{3}$ from the minimal coupling term in equation (\ref{five-dimensional-action}), the minimal coupling contributions lead to $\beta_m^2 F_i^m$ and $\beta_m^2 g_i^m$, which coincide with the results of \cite{Gao:2009ze}, while for $\beta_m^2 F_1^m$ and $\beta_m^2 F_2^m$ we also recover the result of \cite{Polchinski:2002jw}. Also, recall that for $\tau=3$ we have ${\cal {Q}}=\frac{1}{2}$.

\subsection{The Pauli term contributions}

Now, we focus on the contributions from the Pauli interaction term. As mentioned, the angular integrals on $S^5$ lead to certain selection rules. For $\tau=3$ spin-1/2 fermionic operators there are two possibilities in terms of the sign of the mass of each tower  the Kaluza-Klein modes on $S^5$ \cite{Kim:1985ez}.
The dual SYM theory operator of twist 3 and spin 1/2 is ${\cal {O}}_0^{(6)} \sim$tr$(F_+ \lambda_{{\cal {N}}=4})$, with $k=0$. The Kaluza-Klein mass of its dual supergravity field in five dimensions is $m_i=-3/2$ (recall that we set the $S^5$ radius $R=1$).

\subsubsection{Incident state $\lambda_k^-$ and intermediate  state $\lambda_k^-$}

Let us consider the situation where the incident and intermediate states have negative Kaluza-Klein masses $m_i$ and $m_{\cal {X}}$, respectively, which again we express in terms of the $\tilde{m}_i \equiv -m_i$ and $\tilde{m}_{\cal {X}} \equiv -m_{\cal {X}}$ masses defined by the Dirac equation. Considering the limit $M^2 \ll q^2$ we can approximate the solution to first order in $M^2/q^2$, obtaining equation (\ref{lambda0menos}).
This is coupled to a spinor of negative Kaluza-Klein mass
$m_{\cal {X}}$ (but positive $\tilde{m}_{\cal {X}}=-m_{\cal {X}}$) and the mass of the intermediate hadronic state is $M_{\cal {X}}^2=-(P+q)^2$. Since the mass is of the same order as $q$ we cannot approximate it, thus we must use the complete solution of equation (\ref{lambdamenos}).

Let us calculate the relevant contribution from the Pauli interaction term separated in their $\mu$ and $z$ components as follows
\begin{eqnarray}
\bar{\lambda}_{\cal {X}}[\gamma^{\mu},\gamma^{\nu}]\lambda_i&=&
e^{i(P_i-P_{\cal {X}}) \cdot x}  \frac{c_i' c_{\cal {X}}' M_{\cal {X}}^{1/2}}{z_0^{\tilde{m}_i+1}}  z^{\tilde{m}_i+9/2}  \nonumber \\ 
&&\bar{u}_{\sigma {\cal {X}}} \left(  J_{\tilde{m}_{\cal {X}}-\frac{1}{2}}(M_{\cal {X}} z)[\gamma^{\mu},\gamma^{\nu}] P_- \frac{M_i z}{2 \tilde{m}_i+1}+   J_{\tilde{m}_{\cal {X}}-\frac{1}{2}}(M_{\cal {X}} z)[\gamma^{\mu},\gamma^{\nu}]P_+   \right) u_{\sigma i} , \nonumber \\
&& \\
\bar{\lambda}_{\cal {X}}[\gamma^{z},\gamma^{\mu}]\lambda_i&=& e^{i(P_i-P_{\cal {X}}) \cdot x} \frac{c_i' c_{\cal {X}}' M_{\cal {X}}^{1/2}}{z_0^{\tilde{m}_i+1}}   z^{\tilde{m}_i+9/2}\nonumber \\
&&\bar{u}_{\sigma {\cal {X}}} \left(  J_{\tilde{m}_{\cal {X}}-\frac{1}{2}}(M_{\cal {X}} z) \gamma^{\mu}(-2P_+) +  J_{\tilde{m}_{\cal {X}}+\frac{1}{2}}(M_{\cal {X}} z)  \frac{M_i z}{2 \tilde{m}_i +1} \gamma^{\mu} (2P_-)\right)u_{\sigma i} . \nonumber \\
&&
\end{eqnarray}

Now, the corresponding integrals in AdS$_5$ which lead to the matrix elements of the current are\footnote{Notice that as in Section 4.2 we have omitted the constant $\frac{b_{1kk}^{-,-}}{12}$, which implies that the contribution to the structure functions from of this subsection must be multiplied by $\beta_P^2$.}
\begin{eqnarray}
I_1^{P}= \frac{1}{4} \int d^5x \sqrt{-g} \ F_{\mu \nu} \ e^{\mu}_{\hat{\mu}} \ e^{\nu}_{\hat{\nu}} \ \bar{\lambda}_{\cal {X}}^-[\gamma^{\hat{\mu}},\gamma^{\hat{\nu}}]\lambda_i^- \, ,\\
I_2^{P}= \frac{1}{4}  \int d^5x \sqrt{-g} \ F_{z \mu} \ e^{z}_{\hat{z}} \  e^{\mu}_{\hat{\mu}} \ \bar{\lambda}_{\cal {X}}^-[\gamma^{\hat{z}},\gamma^{\hat{\mu}}]\lambda_i^- \, ,
\end{eqnarray}
where the superscript $P$ here stands for the Pauli term.
Thus, we have
\begin{eqnarray}
I_1^P&=&  \frac{1}{4} \int d^5x \ z^{-3} e^{i q \cdot x} q z K_1(q z) i   \left(q_{\mu}n_{\nu}-q_{\nu} n_{\mu}\right) \bar{\lambda}_{\cal {X}}[\gamma^{\mu},\gamma^{\nu}]\lambda_i  \nonumber \\
&=&   \frac{c_i' c_{\cal {X}}' M_{\cal {X}}^{1/2}}{4 z_0^{\tilde{m}_i+1}} \int d^5x e^{i(q + P_i - P_{\cal {X}}) \cdot x} i q K_1(qz) \times \nonumber \\
&& \bar{u}_{\sigma {\cal {X}}}\left( z^{\tilde{m}_i+\frac{7}{2}} J_{\tilde{m}_{\cal {X}}- \frac{1}{2}}(M_{\cal {X}} z)2[\slashed{q},\slashed{n}]\frac{M_i}{2(\tilde{m}_i+\frac{1}{2})}P_- + z^{\tilde{m}_i+\frac{5}{2}} J_{\tilde{m}_{\cal {X}} + \frac{1}{2}}(M_{\cal {X}} z)2[\slashed{q},\slashed{n}] P_+\right)u_{\sigma i} \, ,\nonumber \\
&&
\end{eqnarray}
having used that
\begin{equation}
  \left(q_{\mu}n_{\nu}-q_{\nu} n_{\mu}\right) [\gamma^{\mu}, \gamma^{\nu}]=2[{\slashed{q}},\slashed{n}] \, .
\end{equation}

Next step is to rewrite the integrals in terms of the twist $\tau_i=\tilde{m}_i+3/2$, which leads to
\begin{eqnarray}
I_1^P&=&  \frac{i}{2} \left(\frac{M_{\cal {X}}}{z_0}\right)^{\frac{1}{2}} \left(  \int d^4x \ e^{i(q+P_i-P_{\cal {X}}) \cdot x}\right) \times \nonumber \\
&&   \left[c_i' c_{\cal {X}}' z_0^{-\tau_i+1} \frac{M_i}{2(\tau_i-1)} \left(\int dz K_1(q z)z^{\tau_i+2} J_{\tau_{\cal {X}}-2}(M_{\cal {X}} z) \right)    q \bar{u}_{\sigma {\cal {X}}}[\slashed{q},\slashed{n}] P_- u_{\sigma i}  \right. \nonumber \\
&& \left. + c_i' c_{\cal {X}}' z_0^{-\tau_i+1} \left(\int dz K_1(q z)z^{\tau_i+1} J_{\tau_{\cal {X}}-1}(M_{\cal {X}} z) \right)  q \bar{u}_{\sigma {\cal {X}}}[\slashed{q},\slashed{n}] P_+ u_{\sigma i} \right] \, .
\end{eqnarray}
If we consider the selection rule $\tau_i=\tau_{\cal {X}}$ it reduces to the case discussed by Gao and Mou \cite{Gao:2010qk}.

For the integral $I_2^P$ we may proceed in a similar way
\begin{eqnarray}
I_2^P&=& \frac{1}{4}  \int d^5x \ z^{-3} \ e^{iq \cdot x} \left( -n_{\mu} q^2+(n\cdot q) q_{\mu} \right) z K_0(qz) \bar{\lambda}_{\cal {X}}[\gamma^{z},\gamma^{\mu}]\lambda_i \nonumber \\
&=& \frac{1}{2} \left(\frac{M_{\cal {X}}}{z_0}\right)^{\frac{1}{2}} \left(  \int d^4x e^{i(q+P_i-P_{\cal {X}}) \cdot x}\right) \times \nonumber \\
&& \left[c_i' c_{\cal {X}}' z_0^{-\tau_i+1} \left(- \int dz z^{\tau_i+1} J_{\tau_{\cal {X}}-2}(M_{\cal {X}} z) K_0(qz) \right)\bar{u}_{\sigma {\cal {X}}} \gamma^{\mu} \left((n\cdot q) q_{\mu}-n_{\mu} q^2 \right)P_+u_{\sigma i} \right. \nonumber\\
&&   +\left. \frac{c_i' c_{\cal {X}}' z_0^{-\tau_i+1} M_i}{2(\tau_i-1)}  \left( \int dz z^{\tau_i+2} J_{\tau_{\cal {X}}-1}(M_{\cal {X}} z)K_0(qz) \right) \bar{u}_{\sigma {\cal {X}}} \gamma^{\mu}  \left((n\cdot q) q_{\mu}-n_{\mu} q^2 \right)P_-u_{\sigma i} \right] . \nonumber \\
\end{eqnarray}
Again we consider the selection rule $\tau_i=\tau_{\cal {X}}=\tau$, and obtain the following matrix elements:
\begin{eqnarray}
n_{\mu} \langle J^{\mu} \rangle &=&  \frac{1}{8\pi}\left(\frac{M_{\cal {X}}}{z_0}\right)^{\frac{1}{2}}  \left(i c_{1P} q \bar{u}_{\sigma {\cal {X}}}[\slashed{q},\slashed{n}] P_- u_{\sigma i} + i c_{2P} q  \bar{u}_{\sigma {\cal {X}}}[\slashed{q},\slashed{n}] P_+ u_{\sigma i}\right. \nonumber \\
&&\left. -2 c_{4P} \bar{u}_{\sigma {\cal {X}}} \gamma^{\mu} \left((n\cdot q) q_{\mu}-n_{\mu} q^2 \right)P_+u_{\sigma i}  + 2 c_{3P}  \bar{u}_{\sigma {\cal {X}}} \gamma^{\mu}  \left((n\cdot q) q_{\mu}-n_{\mu} q^2 \right)P_-u_{\sigma i}\right) ,  \nonumber \\
&& \\
n_{\mu} \langle J^{\mu *} \rangle &=& \frac{1}{8\pi}\left(\frac{M_{\cal {X}}}{z_0}\right)^{\frac{1}{2}} \left(-i c_{1P} q \bar{u}_{\sigma i}[\slashed{q},\slashed{n}] P_+ u_{\sigma {\cal {X}}} -i c_{2P} q  \bar{u}_{\sigma i}[\slashed{q},\slashed{n}] P_- u_{\sigma {\cal {X}}} \right. \nonumber \\
&& \left. -2 c_{4P} \bar{u}_{\sigma i} \gamma^{\mu} \left((n\cdot q) q_{\mu}-n_{\mu} q^2 \right)P_+u_{\sigma {\cal {X}}}  + 2              c_{3P}  \bar{u}_{\sigma i} \gamma^{\mu}  \left((n\cdot q) q_{\mu}-n_{\mu} q^2 \right)P_-u_{\sigma {\cal {X}}}\right) ,  \nonumber \\
&&
\end{eqnarray}
where the constants are given by
\begin{eqnarray}
c_{1P}&=&  c_0\frac{M_i}{(\tau-1)}\left(\int dz K_1(q z)z^{\tau+2} J_{\tau-2}(M_{\cal {X}} z) \right) \nonumber \\
%2 c_0 2^{\tau} \frac{M_i}{\tau-1} \tau \Gamma(\tau) M_{\cal {X}}^{-(\tau+4)} q \left(1+\frac{q^2}{M_{\cal {X}}^2} \right)^{-(\tau+2)} \left( \frac{q^2}{M_{\cal {X}}^2}(\tau-1)-2 \right) \nonumber\\
&=&  c_0 2^{\tau+1} \frac{M_i}{\tau-1} \tau \Gamma(\tau) q^{-(\tau+3)} (1-x)^{\frac{(\tau-2                                                      )}{2}} x^{\frac{\tau+4}{2}} \left(x(\tau+1)-2 \right) \, ,\\
c_{2P}&=& 2 c_0 \left(\int dz K_1(q z)z^{\tau+1} J_{\tau-1}(M_{\cal {X}} z) \right)= c_0 2^{\tau+1} \tau \Gamma(\tau) q^{-(\tau+2)} x^{\frac{\tau+3}{2}} (1-x)^{\frac{\tau-1}{2}} \, ,\\
c_{3P}&=& c_0 \frac{M_i }{(\tau-1)} \left( \int dz z^{\tau+2} K_0(qz) J_{\tau-1}(M_{\cal {X}} z) \right) \nonumber \\
&=& c_0 2^{\tau+1} \frac{M_i }{(\tau-1)}  \Gamma(\tau) \tau q^{-(\tau+3)} x^{\frac{\tau+3}{2}} (1-x)^{\frac{\tau-1}{2}} (x(\tau+1)-1) \, , \\
c_{4P}&=&2 c_0  \left(\int dz z^{\tau+1} J_{\tau-2}(M_{\cal {X}} z) K_0(qz) \right) =   c_0 2^{\tau+1}  \Gamma(\tau) q^{-(\tau+2)} x^{\frac{\tau+2}{2}} (1-x)^{\frac{\tau-2}{2}} (x\tau-1) \, , \nonumber \\
\end{eqnarray}
where $c_0=\frac{2 \pi c'_i c'_{\cal {X}}}{z_0^{\tau-1}} K$.
In this case there is an additional factor 2 with respect to the results of Gao and Mou \cite{Gao:2010qk} that we have checked in our calculations. Finally, we obtain the corresponding contributions to the structure functions coming from the Pauli term with no mixing of initial and intermediate states ($\tau_i=\tau_{\cal {X}}=\tau$)
\begin{eqnarray}
F^P_1&=&\frac{F_3^P}{2}=g_1^P=\frac{g_5^P}{2}=\frac{q^6}{8 x}  \left(c_{2P}\left(\frac{1-x}{x}\right)^{1/2}+c_{4P} \right)^2 \nonumber \\
&=&\frac{1}{2} |a_0|^2 \ \Gamma^2(\tau) \left(\frac{\Lambda^2}{q^2}\right)^{\tau-1} x^{\tau+1}(1-x)^{\tau-2} (1-\tau)^2 \, , \\
F^P_2&=&g_4^p=\frac{1}{4} \frac{q^6}{x} \left(c_{2P}^2+c_{4P}^2 \right)= |a_0|^2 \ \Gamma^2(\tau) \left(\frac{\Lambda^2}{q^2}\right)^{\tau-1}  x^{\tau+1}(1-x)^{\tau-2} (1+x\tau(\tau-2)) \, , \nonumber \\
\\
g_2^P&=& - \frac{q^6}{16 x^2} \left( c_{2P}^2 + c_{4P}^2 + 2(c_{2P} c_{1P}+c_{3P} c_{4P})\frac{q}{M} \left(\frac{1-x}{x}\right)^{1/2} \right. \nonumber \\
&& \left. +(c_{1P} c_{4P}-c_{2P}c_{3P}) \frac{q}{M} \frac{2x-1}{x} \right)  \nonumber \\
&=&- \frac{1}{4} |a_0|^2 \ \Gamma^2(\tau) \left(\frac{\Lambda^2}{q^2}\right)^{\tau-1}  (1-x)^{\tau-2} x^{\tau+1}(\tau(1-\tau+x(3+2(\tau-2)\tau))-1)/(\tau-1), \nonumber \\
&& \\
g_3^P&=& \frac{q^6}{4 x}\left(c_{2P}^2+c_{4P}^2-(c_{2P}c_{3P}+c_{1P} c_{4P})\frac{q}{x} \right) \nonumber \\
&=& - |a_0|^2 \ \Gamma^2(\tau) \left(\frac{\Lambda^2}{q^2}\right)^{\tau-1} (1-x)^{\tau-2} x^{\tau+1}(1+\tau (1-3x)+(2x-1)\tau^2)/(\tau-1) \, .
\end{eqnarray}
These functions have several differences compared with the corresponding results of Gao and Mou due to the factor 2 described above.

\subsubsection{Incident state $\lambda_k^-$ and intermediate state $\lambda_{k\pm 1}^+$}

In this case the incident hadron is represented by the dilatino mode given in equation (\ref{lambda0menos}), while the intermediate state corresponds to a Kaluza-Klein mode of positive mass given by equation (\ref{lambdamas}). As before we separate the $\mu$ and $z$ contributions from the Pauli term
\begin{eqnarray}
\bar{\lambda}^+_{\cal {X}}[\gamma^{\mu},\gamma^{\nu}]\lambda^-_i&=&e^{i(P_i-P_{\cal {X}}) \cdot x} \frac{c_i' c_{\cal {X}}' M_{\cal {X}}^{1/2}}{z_0^{\tilde{m}_i+1}}  z^{\tilde{m}_i+9/2} \times \nonumber \\
&&\bar{u}_{\sigma {\cal {X}}} \left(  J_{\tilde{m}_{\cal {X}}+\frac{1}{2}}(M_{\cal {X}} z)[\gamma^{\mu},\gamma^{\nu}] P_- \frac{M_i z}{2 \tilde{m}_i+1}+   J_{\tilde{m}_{\cal {X}}-\frac{1}{2}}(M_{\cal {X}} z)[\gamma^{\mu},\gamma^{\nu}]P_+   \right) u_{\sigma i} , \nonumber \\
&& \\
\bar{\lambda}^+_{\cal {X}}[\gamma^{z},\gamma^{\mu}]\lambda^-_i&=& e^{i(P_i-P_{\cal {X}})\cdot x} \frac{c_i' c_{\cal {X}}' M_{\cal {X}}^{1/2}}{z_0^{\tilde{m}_i+1}} z^{\tilde{m}_i+9/2} \times \nonumber \\
&&\bar{u}_{\sigma {\cal {X}}} \left( J_{\tilde{m}_{\cal {X}}+\frac{1}{2}}(M_{\cal {X}} z) (-2\gamma^{\mu}) P_+ +  J_{\tilde{m}_{\cal {X}}-\frac{1}{2}}(M_{\cal {X}} z)(2\gamma^{\mu})  \frac{M_i z}{2 \tilde{m}_i +1} P_-\right)u_{\sigma i} \, . \nonumber \\
\end{eqnarray}
The integrals are obtained in a similar way as in the previous subsection, but with different relative factors given by different combinations of the Bessel functions\footnote{As in Section 4.3.1 we have omitted the constant $\frac{b_{1jk}^{+,-}}{12}$, which implies that the contribution to the structure functions from this subsection must be multiplied by $\beta_+^2$ for $j=k+1$ and by $\beta_-^2$ for $j=k-1$.},
\begin{eqnarray}
I_1^\pm&=& \frac{i}{2} \left(\frac{M_{\cal {X}}}{z_0}\right)^{\frac{1}{2}} \left(\int d^4x \ e^{i(q+P_i-P_{\cal {X}}) \cdot x}\right) \times \nonumber \\
&&\left[c_i' c_{\cal {X}}' z_0^{-\tau_i+1} \frac{M_i}{2(\tau_i-1)} \left(\int dz K_1(q z)z^{\tau_i+2} J_{\tau_{\cal {X}}-1}(M_{\cal {X}} z) \right)  q \bar{u}_{\sigma {\cal {X}}}[\slashed{q},\slashed{n}] P_- u_{\sigma i}  \right. \nonumber \\
&&  \left. +  c_i' c_{\cal {X}}' z_0^{-\tau_i+1} \left(\int dz K_1(q z)z^{\tau_i+1} J_{\tau_{\cal {X}}-2}(M_{\cal {X}} z) \right)  q \bar{u}_{\sigma {\cal {X}}}[\slashed{q},\slashed{n}] P_+ u_{\sigma i} \right] \, , \\
I_2^\pm&=& \frac{1}{2} \left(\frac{M_{\cal {X}}}{z_0}\right)^{\frac{1}{2}} \left(\int d^4x \ e^{i(q+P_i-P_{\cal {X}}) \cdot x}\right) \times \nonumber \\
&&\left[c_i' c_{\cal {X}}' z_0^{-\tau_i+1} \left(- \int dz z^{\tau_i+1} J_{\tau_{\cal {X}}-1}(M_{\cal {X}} z) K_0(qz) \right)\bar{u}_{\sigma {\cal {X}}} \left((n\cdot q) q_{\mu}-n_{\mu} q^2 \right)P_+u_{\sigma i} \right. \nonumber\\ 
&&   +\left. c_i' c_{\cal {X}}' z_0^{-\tau_i+1}  \left( \int dz z^{\tau_i+2} J_{\tau_{\cal {X}}-2}(M_{\cal {X}} z) K_0(qz) \right) \right. \times \nonumber \\
&&\left.\frac{M_i}{2(\tau_i-1)} \bar{u}_{\sigma {\cal {X}}} \left((n\cdot q) q_{\mu}-n_{\mu} q^2 \right)P_-u_{\sigma i}\right] \, ,
\end{eqnarray}
where the superscript $\pm$ labels the integrals for the present case where the sign of the Kaluza-Klein masses of the initial and the intermediate states are distinct. Then, we calculate the matrix elements of the current 
\begin{eqnarray}
n_\mu \langle J^{\mu} \rangle &=&  \frac{1}{8\pi}\left(\frac{M_{\cal {X}}}{z_0}\right)^{\frac{1}{2}}\left(i c_{1}^{\pm} q \bar{u}_{\sigma {\cal {X}}}[\slashed{q},\slashed{n}] P_- u_{\sigma i} + i c_{2}^{\pm} q  \bar{u}_{\sigma {\cal {X}}}[\slashed{q},\slashed{n}] P_+ u_{\sigma i}\right. \nonumber\\
&& \left.-2 c_{4}^{\pm} \bar{u}_{\sigma {\cal {X}}} \gamma^{\mu} \left((n\cdot q) q_{\mu}-n_{\mu} q^2 \right)P_+u_{\sigma i} + 2 c_{3}^{\pm}  \bar{u}_{\sigma {\cal {X}}} \gamma^{\mu}  \left((n\cdot q) q_{\mu}-n_{\mu} q^2 \right)P_-u_{\sigma i}\right) , \nonumber \\
&& \\
n_\mu \langle J^{\mu*} \rangle &=& \frac{1}{8\pi}\left(\frac{M_{\cal {X}}}{z_0}\right)^{\frac{1}{2}}\left(-i c_{1}^{\pm} q \bar{u}_{\sigma i}[\slashed{q},\slashed{n}] P_- u_{\sigma {\cal {X}}} -i c_{2}^{\pm} q  \bar{u}_{\sigma i}[\slashed{q},\slashed{n}] P_+ u_{\sigma {\cal {X}}}\right. \nonumber\\
&& \left.-2 c_{4}^{\pm} \bar{u}_{\sigma i} \gamma^{\mu} \left((n\cdot q) q_{\mu}-n_{\mu} q^2 \right)P_+u_{\sigma {\cal {X}}} + 2              c_{3}^{\pm}  \bar{u}_{\sigma i} \gamma^{\mu}  \left((n\cdot q) q_{\mu}-n_{\mu} q^2 \right)P_-u_{\sigma {\cal {X}}}\right) . \nonumber \\
\end{eqnarray}
Now, we apply the selection rules we found, namely: $k_{\cal {X}}=k_i\pm1$. With this we obtain the following set of constants: 
\begin{eqnarray}
c_1^+&=& c_0\frac{M_i}{\tau-1} \int dz K_1(q z)z^{\tau+2} J_{\tau+2}(M_{\cal {X}} z) \nonumber \\
&=&
%c_0 2^{\tau} \Gamma(\tau+2) \frac{M_i}{\tau-1} q^{-1}M_{\cal {X}}^{-(\tau+2)} \left( 1+\frac{q^2}{M_{\cal {X}}^2} \right)^{-(\tau+2)}\\&=&
c_0 M_i \frac{\tau+1}{\tau-1} 2^{\tau+1} \Gamma(\tau+1) q^{-(\tau+3)} x^{\frac{\tau+2}{2}} (1-x)^{\frac{\tau+2}{2}} \, ,  \\
c_1^-&=& c_0\frac{M_i}{\tau-1} \int dz K_1(q z)z^{\tau+2} J_{\tau-2}(M_{\cal {X}} z) \nonumber \\
%c_0 2^{\tau} \Gamma(\tau+1) \frac{M_i}{\tau-1} q M_{\cal {X}}^{-(\tau+4)} \left( 1+\frac{q^2}{M_{\cal {X}}^2} \right)^{-(\tau+2)} \left(\frac{q^2}{M_{\cal {X}}^2}(\tau-1)-2\right)\\
%&=&
%
&=&c_0 2^{\tau+1} \Gamma(\tau+1) \frac{M_i}{\tau-1} q^{-(\tau+3)} (1-x)^{\frac{\tau-2}{2}} x^{\frac{\tau+4}{2}} (x(\tau+1)-2) \, ,\\
c_2^+&=&2 c_0\int dz z^{\tau+1} K_{1}(q z) J_{\tau+1}(M_{\cal {X}} z)=
%c_0 2^{\tau} \Gamma(\tau+1) M_{\cal {X}}^{-(\tau+1)} q^{-1} \left(1+\frac{q^2}{M_{\cal {X}}^2}\right)^{-(\tau+1)}\\
%&=&
c_0 2^{\tau+1} \Gamma(\tau+1) q^{-(\tau+2)} x^{\frac{\tau+1}{2}}(1-x)^{\frac{\tau+1}{2}} \, ,\\
c_2^-&=&2 c_0 \int dz z^{\tau+1} K_{1}(q z) J_{\tau-3}(M_{\cal {X}} z) \nonumber \\
&=&% c_0 2^{\tau} \Gamma(\tau) M_{\cal {X}}^{-(\tau+3)}q  \left( 1+\frac{q^2}{M_{\cal {X}}^2} \right)^{-(\tau+1)} (\frac{q^2}{M_{\cal {X}}^2}(\tau-2)-2) \\
%c_2^-&=&
c_0 2^{\tau+1} \Gamma(\tau) q^{-(\tau+2)} x^{\frac{\tau+3}{2}} (1-x)^{\frac{\tau-3}{2}}\left(x\tau-2 \right) \, , \\
c_3^+&=& c_0 \frac{M_i}{\tau-1}\int dz z^{\tau+2} J_{\tau+1}(M_{\cal {X}} z) K_0(qz) \nonumber \\
% c_0 2^{\tau} \frac{M_i}{2(\tau_i-1)} \Gamma(\tau+2) M_{\cal {X}}^{-(\tau+3)}  \left( 1+\frac{q^2}{M_{\cal {X}}^2} \right)^{-(\tau+2)}  \\
%c_2^-&=&
&=&c_0 \frac{M_i}{\tau-1} 2^{\tau+1} q^{-(\tau+3)} \Gamma(\tau+2)x^{\frac{\tau+3}{2}} (1-x)^{\frac{\tau+1}{2}} \, , 
\\
c_3^-&=&c_0 \frac{M_i}{\tau-1}\int dz z^{\tau+2} J_{\tau-3}(M_{\cal {X}} z) K_0(qz) \nonumber \\
% c_0 2^{\tau} \frac{M_i}{(\tau_i-1)}  \Gamma(\tau) M_{\cal {X}}^{-(\tau+3)}  \left( 1+\frac{q^2}{M_{\cal {X}}^2} \right)^{-(\tau+2)} (2-4\frac{q^2}{M_{\cal {X}}^2}(\tau-1)+\frac{q^4}{M_{\cal {X}}^4}(\tau-2)(\tau-1)) \\
%c_2^-&=&
&=& c_0 \frac{M_i}{\tau-1} 2^{\tau+1}  \Gamma(\tau) q^{-(\tau+3)} x^{\frac{\tau+3}{2}} (1-x)^{\frac{\tau+3}{2}}\left( 2+ x \tau(-4+x(\tau+1)  \right) \, ,
\\
c_4^{+}&=&2 c_0 \int dz z^{\tau+1} J_{\tau+2}(M_{\cal {X}} z) K_0(qz) \nonumber \\
% c_0 2^{\tau} \Gamma(\tau+2) M_{\cal {X}}^{(\tau+2)}q^{-2\tau-4} _2F_1(\tau+2,\tau+2,\tau+3,-\frac{M_{\cal {X}}^2}{q^2}) \\
%c_2^-&=&
&=&c_0 2^{\tau+2} \Gamma(\tau+2) q^{-(\tau+2)} x^{-\frac{(\tau+2)}{2}} (1-x)^{\frac{\tau+2}{2}} \phantom F _2 F_1(\tau+2,\tau+2,\tau+3,\frac{x-1}{x}) \, ,
\\
c_4^{-}&=&2 c_0 \int dz z^{\tau+1} J_{\tau-2}(M_{\cal {X}} z) K_0(qz)=% c_0 2^{\tau} \Gamma(\tau) M_{\cal {X}}^{-(\tau+2)}  \left( 1+\frac{q^2}{M_{\cal {X}}^2} \right)^{-(\tau+1)} (\frac{q^2}{M_{\cal {X}}^2}(\tau-1)-1) \\
%c_2^-&=&
c_0 2^{\tau+1} \Gamma(\tau) x^{\frac{\tau+2}{2}} (1-x)^{\frac{\tau-2}{2}}\left(x \tau-1\right) \, ,
\end{eqnarray}
where $\tau_i=\tau$.

Then, for the process $\lambda_k^- + B_\mu^1 \rightarrow \lambda_{k+1}^+$ we obtain the following contributions to the structure functions. The superscript $P_+$ indicates that the intermediate state is $\lambda_{k_{\cal {X}}}^+$, so that $k_{\cal {X}}=k_i + 1$.
\begin{eqnarray}
F^{P+}_1&=&\frac{F_3^{P+}}{2}=g_1^{P+}=\frac{g_5^{P+}}{2}= \frac{q^6}{8 x} \left(c_{2}^+\left(\frac{1-x}{x}\right)^{1/2}-c_{4}^+ \right)^2 \nonumber \\
&=& \frac{1}{2} |a_0|^2 \left(\frac{\Lambda^2}{q^2}\right)^{\tau-1} x^{-\tau-3}(1-x)^{\tau+2} \Gamma(1+\tau)^2 \times \nonumber \\
&& \left[x^{\tau+1}-(1+\tau)\phantom F _2 F_1\left(\tau+2,\tau+2,\tau+3,\frac{x-1}{x}\right) \right]^2 \, , \\
F^{P+}_2&=&g_4^{P+}= \frac{q^4}{16 x} \left((c_{2}^+)^2+(c_{4}^+)^2 \right) = |a_0|^2 \left(\frac{\Lambda^2}{q^2}\right)^{\tau-1} x^{-\tau}(1-x)^{\tau+1} \Gamma(1+\tau)^2 \times \nonumber \\
&&\left[x^{2\tau}+ \frac{1-x}{x^3}(1+\tau)^2 \phantom F _2 F_1\left(\tau+2,\tau+2,\tau+3,\frac{x-1}{x} \right)^2 \right] \, , \\
g_2^{P+}&=& - \frac{q^6}{16 x^2} \left( (c_{2}^+)^2 + (c_{4}^+)^2 - 2(c_{2}^+ c_{1}^+ +c_{3}^+ c_{4}^+)\frac{q}{M} \left(\frac{1-x}{x}\right)^{1/2}+(c_{1}^+ c_{4}^+-c_{2}^+ c_{3}^+) \frac{q}{M} \frac{2x-1}{x} \right) \nonumber \\
&=&- \frac{1}{4} |a_0|^2 \left(\frac{\Lambda^2}{q^2}\right)^{\tau-1} (1-x)^{\tau} x^{\tau} \Gamma(1+\tau)^2 \times \nonumber \\
&&\left[ 2\frac{1-x}{1-\tau}+ (1-x)^{-2\tau} (1+\tau)^2 (2+\tau) B(\frac{x-1}{x},\tau+2,-(\tau+1)) \right. \nonumber \\
&& \left. \left( \frac{(x-1)^ {\tau}}{1-\tau}+\frac{x(\tau+2) B(\frac{x-1}{x},\tau+2,-(\tau+1))}{(1-x)^2} \right)\right] \, , \\
g_3^{P+}&=&\frac{q^6}{4 x}\left((c_{2}^+)^2+(c_{4}^+)^2-(c_{2}^+c_{3}^++c_{1}^+ c_{4}^+)\frac{q}{x} \right) \nonumber \\
&=& |a_0|^2 \left(\frac{\Lambda^2}{q^2}\right)^{\tau-1} \frac{1}{\tau-1} 8 (1-x)^{\tau+1} x^{-\tau-3} \Gamma(1+\tau)^2 \times \nonumber \\
&& \left[ 2x^{2\tau+3}+ (1+\tau)^2(x-1) \phantom F _2 F_1(\tau+2,\tau+2,\tau+3,\frac{x-1}{x}) \right. \nonumber \\
&&  \left. \left( -x^{\tau+1}+ (\tau-1) \phantom F _2 F_1(\tau+2,\tau+2,\tau+3,\frac{x-1}{x})\right) \right] \, ,
\end{eqnarray}
where $_2F_1$ is the hypergeometric function and $B(x,a,b)$ is the incomplete Beta function.

Analogously, for the process $\lambda_k^- + B_\mu^1 \rightarrow \lambda_{k-1}^+$ we obtain the following contributions to the structure functions. The superscript $P_-$ indicates that the intermediate state is $\lambda_{k_{\cal {X}}}^+$, now being $k_{\cal {X}}=k_i - 1$.
\begin{eqnarray}
F^{P-}_1&=&\frac{F_3^{P-}}{2}=g_1^{P-}=\frac{g_5^{P-}}{2}=\frac{q^6}{8 x} \left(c_{2}^-\left(\frac{1-x}{x}\right)^{1/2}-c_{4}^- \right)^2 \nonumber \\
&=& \frac{1}{2} |a_0|^2 \left(\frac{\Lambda^2}{q^2}\right)^{\tau-1}(1-x)^{\tau-2} x^{\tau+1} \Gamma(\tau)^2 \, , \\
F^{P-}_2&=&g_4^{P-}=\frac{q^4}{16 x} \left((c_{2}^-)^2+(c_{4}^-)^2 \right) \nonumber \\
&=& |a_0|^2 \left(\frac{\Lambda^2}{q^2}\right)^{\tau-1}(1-x)^{\tau-3} x^{\tau+1} \Gamma(\tau)^2(1+x(3+\tau(-2+x(\tau-2)))) \, ,\\
g_2^{P-}&=&\frac{-q^6}{16 x^2} \left( (c_{2}^-)^2 + (c_{4}^-)^2 - (c_{2}^- c_{1}^-+c_{3}^- c_{4}^-)\frac{2q}{M} \left(\frac{1-x}{x}\right)^{1/2}+(c_{1}^- c_{4}^--c_{2}^-c_{3}^-) \frac{q}{M} \frac{2x-1}{x} \right) \nonumber \\
&=& -\frac{1}{4} |a_0|^2 \left(\frac{\Lambda^2}{q^2}\right)^{\tau-1}\frac{\Gamma(\tau)^2}{\tau-1} x^{\tau+1} (1-x)^{\tau-3} \left[- \tau-1+x\left(x^{5}(\tau+1)\tau (\tau-2)+(1+\tau)^2 \right.\right. \nonumber \\
&& \left.\left. -x^4(\tau-2)\tau (7+3\tau)+6x^2(2+(3-2\tau)\tau) \right. \right. \nonumber\\
&&  \left. \left. -x (12+(\tau-5)\tau)+x^3(\tau-3)(2+3\tau(5+\tau)) \right) \right] \, ,\\
g_3^{P-}&=&\frac{q^6}{4 x}\left((c_{2}^-)^2+(c_{4}^-)^2-(c_{2}^-c_{3}^-+c_{1}^- c_{4}^-)\frac{q}{x} \right) 
= |a_0|^2 \left(\frac{\Lambda^2}{q^2}\right)^{\tau-1} \frac{\Gamma(\tau)^2}{\tau-1} x^{\tau+1}(1-x)^{\tau-3} \times \nonumber \\
&&\left( 3-\tau+x\left( -15-4(x-3)x-2\tau+x(33-2x(18+(x-8)x))\tau   \right. \right. \nonumber \\
&&  \left. \left.+(1+x(-1+(x-6)x(3+x(x-3))))\tau^2+x^3 (3+x(x-3)\tau^3)  \right) \right) \, .
\end{eqnarray}
We should notice that all these contributions $F_i^{P-}$'s and $g_i^{P-}$'s are not present when $\tau=3$ since in that case the label $k$ of the incident state is zero, therefore the intermediate state can only have $j=k+1$. This overcomes the fact that $F_2^{P-}$, $g_2^{P-}$ and $g_3^{P-}$ behave like 
$(1-x)^{\tau-3}$.

\subsection{The mixed contribution from minimal coupling and Pauli vertices}
 
In addition to the set of contributions we have already calculated, there is a mixed contribution corresponding to a Feynman diagram for the forward Compton scattering in which there is a minimal coupling vertex and a Pauli vertex. This leads to matrix elements of the hadronic tensor of the form
\begin{equation}
    n_{\mu} n_{\nu}   W^{c\mu \nu}=n_{\mu} n_{\nu} \ 2 \pi^2 \sum_{{\cal {X}}}\delta(M_{\cal {X}}^2+(P+q)^2) (\langle J_m^{\mu}(0)\rangle \langle J_P^{*\nu}(0)\rangle+\langle J_P^{\mu}(0)\rangle \langle J_m^{*\nu}(0)\rangle) \, ,
\end{equation}
where the matrix elements of the currents correspond to contributions from minimal coupling or Pauli interaction term.
The superscript $c$ indicates crossed-term contributions. The structure functions from crossed terms are:
\begin{eqnarray}
F^{c}_1&=&\frac{F_3^{c}}{2}=g_1^c=\frac{g_5^c}{2}= \frac{q^5}{2 x^2} c_{1m} \left(c_{2P}\left(\frac{1-x}{x} \right)^{1/2} +c_{4P} \right) \nonumber \\
&=& \frac{1}{2} |a_0|^2 \left(\frac{\Lambda^2}{q^2}\right)^{\tau-1}  (1 - x)^{\tau-2} x^{\tau+1} (\tau-1) \Gamma(\tau)^2  \, , \\
F^{c}_2&=&g_4^c= 2 \frac{q^5}{x} c_{1m}c_{4P} \nonumber
\\
&=& |a_0|^2 \left(\frac{\Lambda^2}{q^2}\right)^{\tau-1} (1 - x)^{\tau-2} x^{\tau+1} (-1 + x\tau) \Gamma(\tau)^2 \, ,
\\
g_2^{c}&=&- \frac{1}{4} \left(q(\frac{1-x}{x})^{1/2} (c_{2P}c_{3m}+c_{3P} c_{1m}-c_{1P} c_{4m}-c_{4P}c_{2m})+(c_{1m} c_{4P}-c_{2P} c_{4m}) \right. \nonumber \\
&&  \left. + q(c_{4m} c_{3P}+c_{4P}c_{3m}+c_{1P}c_{1m}+c_{2P} c_{2m})  \right) \nonumber \\
&=&- \frac{1}{4} |a_0|^2 \left(\frac{\Lambda^2}{q^2}\right)^{\tau-1}\frac{\Gamma(\tau)^2}{\tau-1} (1-x)^{\tau-2} x^{\tau+2} (2-\tau^2+x\tau(4\tau-5)) \, ,
\\
g_3^{c}&=& 2 \frac{q^5}{x} \left( c_{1m}c_{4P}+(c_{1m} c_{1P}-c_{2m} c_{2P})\frac{q}{2x}  \right) \nonumber \\
&=&\frac{1}{2} \frac{\Gamma(\tau)^2}{\tau-1} |a_0|^2 \left(\frac{\Lambda^2}{q^2}\right)^{\tau-1} (1-x)^{\tau-2} x^{\tau+1} \left(2-(4+x)\tau +(-1+4x)\tau^4 \right) \, .
\end{eqnarray}

\section{Results of the structure functions from the $\tau=3$ spin-1/2 fermionic operator}

In the previous section we have obtained the contributions coming form each Feynman diagram to the structure functions related to spin-1/2 fermionic operators. The complete expression for each structure function can be written as the sum of those contributions with relative constants multiplying each diagram.
The referred constants are given in terms of certain angular integrals that we explicitly calculate for $\tau=3$. Thus, the general form of the structure functions is
\begin{equation}
    F_i= \beta^2_m F_i^m + \beta^2_P F_i^P + \beta_m \beta_P F_i^c+\beta^2_+ F_i^{P+} + \beta^2_- F_i^{P-} \, ,
    \label{betas}
\end{equation}
and there is a similar expression for the $g_i$ structure functions. The $\beta$ constants in equation (\ref{betas}) are straightfordwarly related to the $b_{1kj}$ angular integrals, being their explict relations given in Appendix A. $\beta_m$ is the constant related to the minimal coupling, being $\beta_m=\frac{{\cal{Q}}}{3}$. Then, $\beta_P$ is the constant associated with Pauli interaction diagrams with the selection rule $\tau_i=\tau_{\cal {X}}$, {\it i.e.} with no mixing of states between the incident and the intermediate hadrons. In addition, in terms of the optical theorem there is also an $s$-channel Feynman diagram containing both the minimal coupling vertex and the Pauli interaction vertex. This is associated with the third contribution $F_i^c$ that we call crossed-terms contribution. The relative constant associated with this contribution is given by the product of $\beta_m$ and $\beta_P$. In addition, $\beta_+$ and $\beta_-$ correspond to relative constants of Feynman diagrams with Pauli interactions with incoming states dual to the operator ${\cal{O}}_k^{(6)}$ but intermediate states dual to ${\cal{O}}_{k'}^{(13)}$ with $k'=k+1$ and $k'=k-1$, respectively. These contributions that couple fermionic modes of the two different Kaluza-Klein towers of the type IIB supergravity compactified on $S^5$ have not been studied in previous papers on DIS. The detailed calculation for general twist operators needs the construction of general spinor spherical harmonics on $S^5$. Since in this work we focus on the case of $\tau=3$, we leave the general case to be discussed in a forthcoming work \cite{Newpaper}. However, by adding all the mentioned contributions to each structure function one obtains the following relations which hold for any twist:
\begin{eqnarray}
F_1&=&\frac{F_3}{2}=g_1=\frac{g_5}{2} \, , \label{FFGG} \\
F_2&=&g_4 \, . \label{FG}
\end{eqnarray}
This is an important general twist result of the present work which is valid in the regime $\lambda_{SYM}^{-1/2} \ll x < 1$. Also, it is interesting to emphasize that the structure functions $F_1$ and $F_2$ above, which include all possible contributions discussed in this work do not satisfy the same relation as in the case considered by Polchinski and Strassler ($F_2=2 F_1$), where they have only included the minimal coupling. This is a consistent result, since from hadron phenomenology it is expected a non-vanishing longitudinal structure function. Also, there are differences with respect to \cite{Gao:2009ze} and \cite{Gao:2010qk}. In \cite{Gao:2009ze} they only included the minimal coupling interaction, while in \cite{Gao:2010qk} only the Pauli term was considered. As already commented, we have done a fully consistent derivation from type IIB supergravity, thus the relations (\ref{FFGG}) and (\ref{FG}) should be regarded as the complete set of Callan-Gross type of relations for all spin-1/2 hadrons related to the ${\cal {O}}_k^{(6)}$ operators of the planar limit of the strongly coupled ${\cal {N}}=4$ SYM theory within $\lambda_{SYM}^{-1/2} \ll x < 1$.

Let us now discuss more explicitly the case with $\tau=3$, which leads to the most important contribution to the Operator Product Expansion (OPE) of two electromagnetic currents inside the hadron at strong 't Hooft coupling and large $N$ ($1 \ll \lambda_{SYM} \ll N$) in the DIS limit, since this is the lowest twist operator in that parametric regime. The corresponding SYM theory operator is ${\cal{O}}^{(6)}_{k=0}$
which has the minimal conformal dimension $\Delta=\frac{7}{2}$.
The dual supergravity field mode has the quantum numbers of its spinor spherical harmonic $(l_5,l_4,l_3,l_2,l_1)=(0,0,0,0,0)$. In particular, the identification $k=l_5=0$ implies that $\tau=3$, while $l_1=0$ leads to ${\cal{Q}}=\frac{1}{2}$, being ${\cal{Q}}$ defined in equation (\ref{eigenvalueeqcharge}).

For $\tau=3$ the constants in equation (\ref{betas}) can be explicitly calculated, obtaining
\begin{eqnarray}
\beta_m=1/6 \ , \ \ \ \ \beta_P=-5/36 \ , \ \ \ \ (\beta_+)^2= 1/648 \ , \ \ \ (\beta_-)^2=0 \, .
\end{eqnarray}

Now, we can draw the corresponding structure functions for $\tau=3$. In each figure we describe the different contributions to the structure functions from each term. 
\begin{figure}[H]
\centering
%\begin{subfigure}[]{
\includegraphics[scale=1.2]{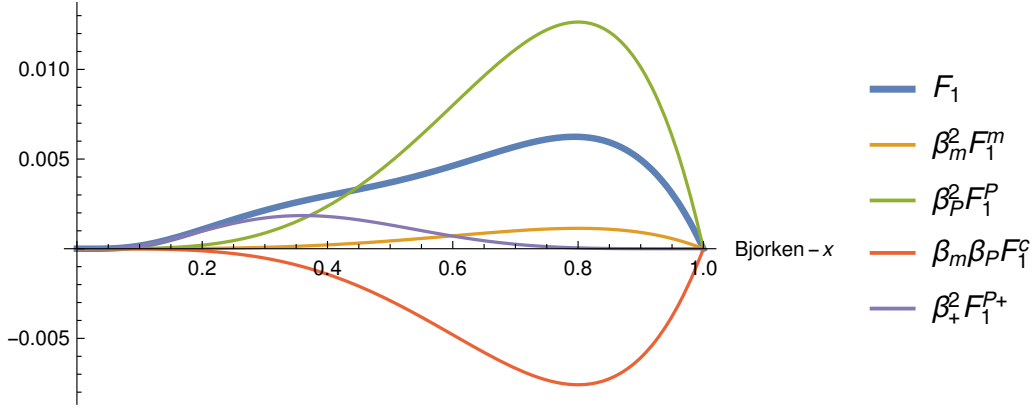}
\label{F1tau3}
\caption{\small The full structure function $F_1$ (blue line) as a function of the Bjorken parameter $x$, obtained from the contribution of the twist-3 spin-1/2 fermionic operator ${\cal {O}}_{k=0}^{(6)}$. We distinguish the contributions from the minimal coupling $\beta^2_m F_1^m$ (orange line); the Pauli interaction $\beta^2_P F_1^P$, where the intermediate state $\lambda_{\cal {X}} \equiv \lambda_{k=0}^-$ is the same as the incident state $\lambda_i  \equiv \lambda_{k=0}^-$ (green line); the contribution from crossed terms $\beta_m \beta_P F_1^c$ (red line); and the contribution from the Pauli interaction $\beta^2_+ F_1^{P +}$, where the intermediate state $\lambda_{\cal {X}} \equiv \lambda_{k+1}^+$ is different from the incident state $\lambda_{\cal {X}} \equiv \lambda_k^-$ according to the selection rules that we found (violet line). We have set $|a_0|=1$ which is the only free constant for all the structure functions.}
\end{figure}
\begin{figure}[H]
\centering
%\begin{subfigure}[]{
\includegraphics[scale=1.2]{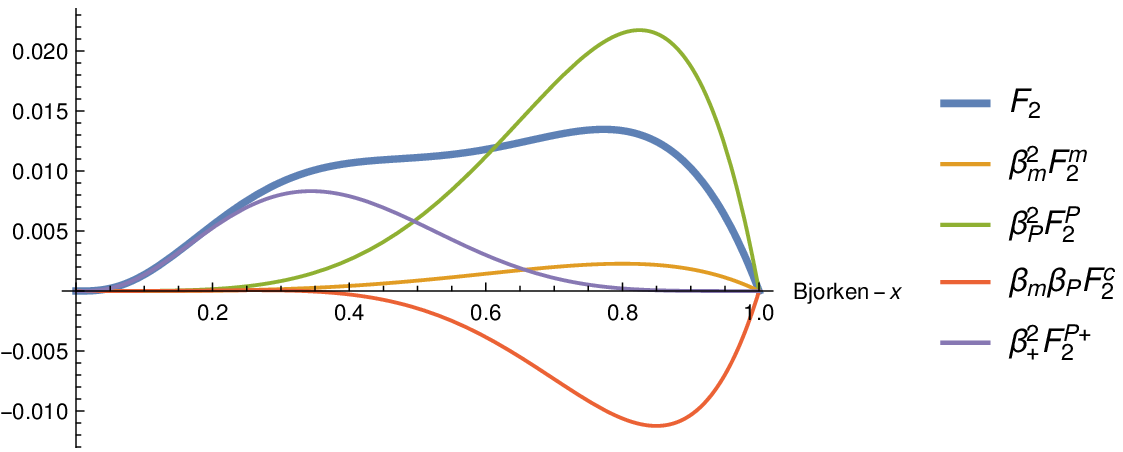}
\label{F2tau3}
\caption{\small The full structure function $F_2$ (blue line) as a function of the Bjorken parameter $x$, obtained from the contribution of the twist-3 spin-1/2 fermionic operator ${\cal {O}}_{k=0}^{(6)}$. The meaning of the curves is analogous as describe in figure 3.}
\end{figure}
\begin{figure}[H]
\centering
%\begin{subfigure}[]{
\includegraphics[scale=1.2]{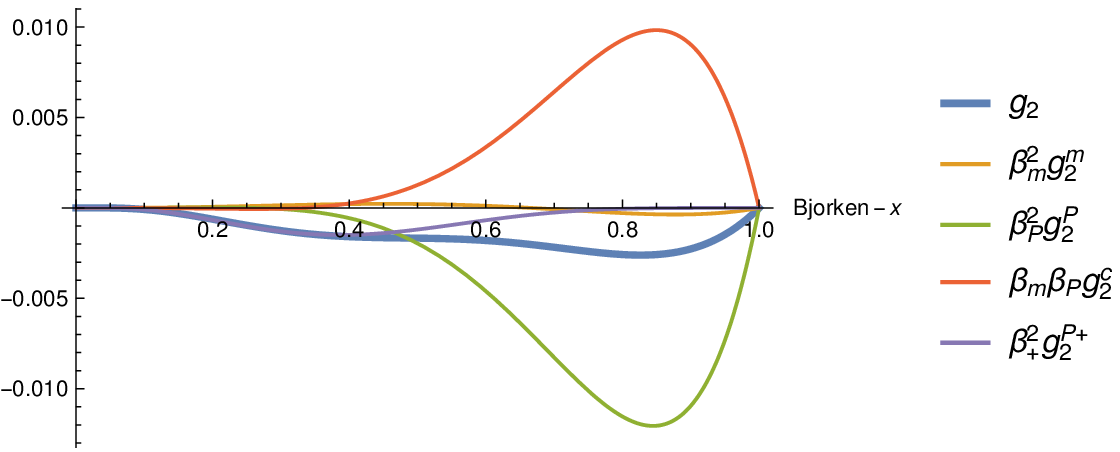}
\label{g2tau3}
\caption{\small The full structure function $g_2$ (blue line) as a function of the Bjorken parameter $x$, obtained from the contribution of the twist-3 spin-1/2 fermionic operator ${\cal {O}}_{k=0}^{(6)}$. The meaning of the curves is analogous as describe in figure 3.}
\end{figure}
\begin{figure}[H]
\centering
%\begin{subfigure}[]{
\includegraphics[scale=1.2]{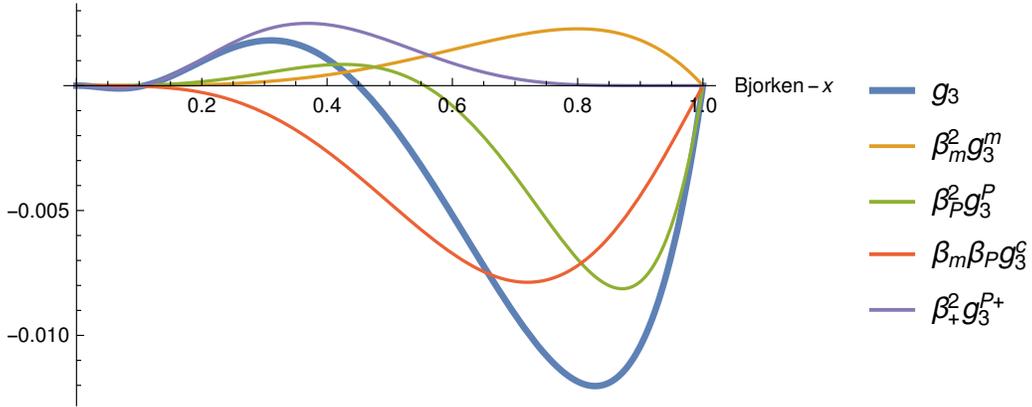}
\label{g3tau3}
\caption{\small The full structure function $g_3$ (blue line) as a function of the Bjorken parameter $x$, obtained from the contribution of the twist-3 spin-1/2 fermionic operator ${\cal {O}}_{k=0}^{(6)}$. The meaning of the curves is analogous as describe in figure 3.}
\end{figure}

In this case the diagram connecting an initial state with $k_i=l_5$ with an intermediate state with $k_{\cal {X}}=l_5-1$ is not present since $l_5=0$. We have found a very interesting result, namely: for all the structure functions the contributions from the minimal coupling (orange curves) are very small in comparison with the rest of contributions. Also, in table 1 we can appreciate the numerical value of the maximum (or minimun value whenever it corresponds) of each curve, which shows that the minimal coupling contribution is very small in comparison with the rest of contributions.  Indeed, within the Bjorken parameter range $0.4 \leq x \leq 1$ the dominant contribution comes from the Pauli term (green curves), followed by the crossed-terms contributions (red curves) which appear with opposite sign (except for $g_3$). Surprisingly, the minimal coupling contributions are really much less significant in comparison with the previously mentioned contributions as well as the total contribution (blue curves). By virtue of the relations (\ref{FFGG}) and (\ref{FG}), this holds for all structure functions. This implies that the theoretical calculation from first principles of all contributions to the forward Compton scattering as well as the relative constants $\beta$'s that we have performed in the present work are very important in order to give the precise contribution of each term.

Thus, we observe that the dominant contribution for $F_1$, $F_2$ and $g_2$ comes from the Feynman diagram of figure 2 with the Pauli vertex (the dotted vertex in the first line), being the outgoing state identical to the incoming one. However, the crossed term in the forward Compton scattering Feynman diagram produces a certain suppression. On the other hand, in the case of $g_3$ the crossed term gives a contribution which enhances the contribution from the Pauli term. This behaviour is different in comparison with the other structure functions.

The contribution of the Feynman diagrams which couple to the other tower of Kaluza-Klein modes with positive five-dimensional masses becomes significant for smaller values of the Bjorken parameter, having a pick around $x=0.35$. For larger values of $x$ its contribution becomes much smaller than the Pauli one.

%%%%
\begin{table}

\def\arraystretch{1.5}
\begin{center}
\begin{tabular}{|c|c|c|c|c|c|c|c|}
\hline
$F_1$ terms & M.V. & $F_2$ terms & M.V. &$g_2$ terms & M.V. &$g_3$ terms & M.V. \\
\hline
$F_{1}^m$ & 0.0011  &  $F_{2}^m$  & 0.0022&$g_{2}^m$  &-0.0003 &$g_{3}^m$  &0.0025  \\
\hline
$F_{1}^P$ & 0.0126 & $F_{2}^P$  &0.02173 &$g_{2}^P$  &-0.0120 &$g_{3}^P$  & -0.0081 \\
\hline
$F_{1}^c$ & -0.0075& $F_{2}^c$ &-0.0112 &$g_{2}^c$  &0.0098 &$g_{3}^c$  & -0.0078 \\
\hline
$F_{1}^{P+}$ & 0.0018 &  $F_{2}^{P+}$  &0.0083 &$g_{2}^{P+}$ &-0.0014 &$g_{3}^{P+}$  & 0.0025 \\
\hline
\end{tabular}
\caption{\small Maximum values (M.V.) (or minimum values wherever there is a negative sign) of the curves depited in figures 3, 4, 5 and 6. We have set $|a_0|=1$.}\label{table}
\end{center}

\end{table}

\section{Discussion and conclusions}

In this work we have investigated the polarised deep inelastic scattering of charged leptons off spin-1/2 hadrons in the ${\cal{N}}=4$ SYM theory deformed by the introduction of the IR scale $\Lambda$. Using the gauge/gravity duality we have calculated the structure functions in the large $N$ limit, and at strong coupling, focusing on the contribution of the leading twist operator. The analysis has been carried out in the Bjorken parameter range $\lambda_{SYM}^{-1/2}\ll x<1 $, where we can use the type IIB supergravity description. It is important to emphasize that our results are derived from first principles (top-down approach) and it implies considering the complete supergravity interactions at the leading order in the $\frac{1}{N}$ expansion (tree-level Witten diagrams), which ultimately can be obtained from the type IIB superstring theory in AdS$_5\times S^5$ in the $\alpha' \rightarrow 0$ limit. These results contrast with the previously calculated ones, which were obtained using either only the minimal coupling interaction (within a top-down approach) \cite{Polchinski:2002jw} or including the Pauli interaction in the framework of bottom-up models \cite{Gao:2009ze,Gao:2010qk}.  

In order to perform the analysis by considering the complete set of interactions, we need to obtain the interaction vertices from the ten-dimensional type IIB supergravity action. However, since the five-form field strength obeys a self-duality constraint in type IIB supergravity, there is no simple covariant action and the constraint needs to be imposed after deriving the equations of motion. For this reason, we have focused on the covariant equations of motion and expanded the ten-dimensional fields in their Kaluza-Klein modes. Thus, we have derived the field equations up to second order in the corresponding fluctuations. Finally, we have constructed the five-dimensional effective action up to cubic order in the fields with dilatino modes and the massless gauge field. The interaction terms are the minimal coupling term and the Pauli term, with relative constants between both interactions which depend on the angular integrals of spinor spherical harmonics. 
The derivation we have done for the interaction terms of the spin-1/2 fermionic field modes with a massless gauge field has not been carried out previously. Thus, it opens several potential interesting directions to further explore. For instance, it would be very interesting to use this action to obtain the three-point functions on ${\cal {N}}=4$ SYM theory involving fermionic operators. Also, using an analogous approach as considered in this work it would be interesting to unveil the effects of the new terms of the effective action on Drell-Yang processes, form factors and other observables.

At this point we should emphasize the importance of having developed a first-principle derivation of the effective five-dimensional supergravity action directly from type IIB supergravity. First, it allows us to find all the consistent interactions involving two dilatini with a massless vector field mode. Then, by constructing the corresponding spinor spherical harmonics followed by solving the corresponding angular integrals on $S^5$, we discover new selection rules between incident and intermediate fermionic states. Moreover, these integrals give the precise values of all the relative constants in front of each contribution, allowing us to calculate the complete set of structure functions from the analysis of the dual twist-3 spin-1/2 operator. The selection rules have been obtained from the evaluation of the spinor spherical harmonics integrals and we have found, that in adittion to the minimal coupling term and the Pauli term studied in \cite{Polchinski:2002jw,Gao:2010qk}, there is a new interaction between states in different irreducible representations of $SU(4)$ (the other Kaluza-Klein tower with $\lambda_{k+1}^+$ modes) and different twist. Let us emphasize that these vertices are new. For $F_1$ and $F_2$, these interactions are responsible for the bell-shaped curves with a maximum at $x \sim 0.35$. Notice that their maximum values occur at smaller values of $x$ in comparison with the others contributions.

Finally, we have obtained the independent structure function $F_1$, $g_2$, $g_3$ and $F_2$, while the rest are related to them through the relations  (\ref{FFGG}) and (\ref{FG}). The relation  $F_2=2 F_1$ obtained in \cite{Polchinski:2002jw} does not hold due to the contributions from the Pauli terms. The relative constants $\beta$'s show that the minimal coupling contributions are very small compared with the Pauli interaction (less than 10$\%$). For $x$ values in the range $0.4<x< 1$ the dominant contribution corresponds to the Pauli term with the same final state. However, the crossed term, which leads to contributions with a different sign, attenuate the final result. On the other hand, within the parametric range $0.2 < x < 0.4$ the dominant contribution comes from the Pauli interaction connecting different states $\lambda_k^-$ and $\lambda_{k+1}^+$.
This is also a novel effect. \\

~

%\newpage

%%%%%%%%%%%%%%%%%%%%%%%%%%%%%%%%%%%%%%%%%%
%
\centerline{\large{\bf Acknowledgments}}
%
%%%%%%%%%%%%%%%%%%%%%%%%%%%%%%%%%%%%%%%%%%

~

This work has been supported by the National Scientific Research Council of Argentina (CONICET), the National Agency for the Promotion of Science and Technology of Argentina (ANPCyT-FONCyT) Grants PICT-2015-1525 and PICT-2017-1647, the UNLP Grant PID-X791 and the CONICET Grant PIP-UE B\'usqueda de nueva f\'{\i}sica.

\newpage

\appendix

\section{Appendix: Angular integrals}

We show how to simplify the integral $b_{1 k j}$ using properties of the spinor spherical harmonics. The spinor spherical harmonics satisfy the Dirac equation being charge eigenstates with angular momentum associated to the angle $\theta_1$, 
\begin{eqnarray}
\tau^{\alpha} \nabla_{\alpha}\Theta_k^{\pm}&=& \mp i\left(k+\frac{5}{2}\right) \Theta_k^{\pm} \label{harm-dirac-1} \, , \\
\left( v^{\alpha} D_{\alpha}-\frac{1}{4}\tau^{\alpha} \tau^{\gamma} \nabla_{\gamma}v_{\alpha} \right) \Theta^{\pm}_{k} &=&-i{\cal{Q}} \Theta^{\pm}_k \, .\label{harm-carga-1}
\end{eqnarray}
The spinor spherical harmonics with positive eigenvalues in equation (\ref{harm-dirac-1}) correspond to the representations {\bf 4$^*$}, {\bf 20$^*$}, {\bf 60$^*$}, $\cdots$, while those with negative eigenvalues belong to the conjugate representations {\bf 4}, {\bf 20}, {\bf 60}, $\cdots$. In the DIS process we consider the holographic dual incident hadron represented by a dilatino field mode containing a spinor spherical harmonic of the type $\Theta_k^-$. Recall that when $k=0$ the dual ${\cal {N}}=4$ SYM theory operator has twist $\tau=3$. 

The second term in equation (\ref{coeff-b}) can be written in terms of the first integral, thus  $b_{1 k j}$ can be expressed by a single angular integral: 

\begin{eqnarray}
\int d \Omega_5  (\Theta_{j}^{\pm})^{ \dag}\left(i \tau^{\alpha}\tau^{\beta} \nabla_{\alpha}v_{\beta} \right)\Theta^-_{k} 
&=& -i \int d \Omega_5 \left(D_{\alpha}(\Theta^{\pm}_{j})^{\dag} \tau^{\alpha}\tau^{\beta} v_{\beta} \Theta^-_{k} + (\Theta^{\pm}_{j})^{\dag} \tau^{\alpha}\tau^{\beta} v_{\beta} D_{\alpha}\Theta^-_{k}\right)\nonumber \\
&=&-i \int d \Omega_5 \left(D_{\alpha}(\Theta^{\pm}_{j})^{\dag} \tau^{\alpha}\tau^{\beta} v_{\beta} \Theta^-_{k} + (\Theta^{\pm}_{j})^{\dag} 2 \eta^{\alpha \beta} v_{\beta} D_{\alpha}\Theta^-_{k}\right.\nonumber \\
&&   -\left. (\Theta^{\pm}_{j})^{\dag} \tau^{\beta} v_{\beta} \tau^{\alpha} D_{\alpha}\Theta^-_{k}\right) \nonumber\\
&=& \left(\pm \left(j+\frac{5}{2}\right)-\left(k+\frac{5}{2}\right)\right) \int   d \Omega_5 (\Theta^{\pm}_{j})^{\dag}  \tau^{\beta} v_{\beta} \Theta_{k}^-  \nonumber \\
&&  -2 {\cal{Q}} \int   d \Omega_5 (\Theta^{\pm}_{j})^{\dag}   \Theta^-_{k}  \nonumber \\
&& -2i \int d \Omega_5  (\Theta^{\pm}_{j})^{\dag} \left(\frac{1}{4} \tau^{\beta}\tau^{\alpha} \nabla_{\alpha}v_{\beta} \right)\Theta^-_{k} \, .
\end{eqnarray}
In order to simplify the integral we use equations (\ref{harm-dirac-1}) and (\ref{harm-carga-1}), obtaining
\begin{eqnarray}
\int d \Omega_5  (\Theta^{\pm}_{j})^{\dag}\left(i \tau^{\alpha}\tau^{\beta} \nabla_{\alpha}v_{\beta} \right)\Theta_{k}^- 
&=& \left(\pm j-k\pm\frac{5}{2}-\frac{5}{2}\right)2  \int   d \Omega_5 (\Theta^{\pm}_{j})^{\dag} \tau^{\beta} v_{\beta} \Theta^-_{k}\nonumber \\
&& \ \  
-4   {\cal{Q}} \int   d \Omega_5 (\Theta^{\pm}_{j})^{\dag}  \Theta^-_{k}  \, .
\end{eqnarray}
Then, the constant $b_{1 k j}$ can  written in terms of a single angular integral
\begin{eqnarray}
b^{\pm,-}_{1 k j}&=&\left(1+ 2 \left(k\mp j+\frac{5}{2}\mp\frac{5}{2}\right) \right) \int d \Omega_5  (\Theta^{\pm}_{j})^{\dag} \tau_{\alpha}v^{\alpha}\Theta_{k} +4{\cal{Q}} \  \delta_{j k}^{\mp,-} \, .
\end{eqnarray}
The second term is present if the intermediate state is the same as the incident one. The constants associated with operators in the same representation are
\begin{eqnarray}
b^{-,-}_{1 k j}&=&\left(1+2(k+j+5) \right)\int d \Omega_5  (\Theta^{-}_{j})^{\dag} \tau_{\alpha}v^{\alpha}\Theta_{k}^- +4 {\cal{Q}} \  \delta_{j k} \, .
\end{eqnarray}
For the interactions with supergravity field modes belonging to the ${\bf 4}$, ${\bf 20}$, ${\bf 60}$, $\dots$ representations the term proportional to the Kronecker delta is not present and we obtain
\begin{eqnarray}
b^{-,+}_{1 k j}&=&\left(1+2(k-j)  \right)\int d \Omega_5  (\Theta^{+}_{j})^{\dag} \tau_{\alpha}v^{\alpha}\Theta_{k}^- \, .
\end{eqnarray}

In section 5 we consider the minimal $\tau=3$ which corresponds to $k=0$ for the incident fermion. We define the following coupling constants in order to draw the structure functions in figures 3, 4, 5 and 6,
\begin{eqnarray}
\beta_P &=& \frac{b_{1 \ 0 \ 0}^{-,-}}{12}=\frac{11\int d \Omega_5  (\Theta^{-}_{0})^{\dag} \tau_{\alpha}v^{\alpha}\Theta_{0}^- +2}{12} \, , \\
(\beta_P^+)^2&=&\sum_{I_5} \left(\frac{b^{-,+}_{1 \ 0 \ 1}}{12}\right)^2=\sum_{I_5} \left(-\frac{\int d \Omega_5  (\Theta^{+}_{1})^{\dag} \tau_{\alpha}v^{\alpha}\Theta_{0}^-}{12} \, \right)^2 \, ,
\end{eqnarray}
where the sum over $I_5$ indicates the sum over the angular integrals given in equations (\ref{integral-1}), (\ref{integral-2}), (\ref{integral-3}) and (\ref{integral-4}).

\section{Appendix: Details of the calculation of the minimal coupling contributions}

In order to calculate the relevant contributions from the minimal coupling term we explicitly write down the $\mu$ and $z$ components of the five-dimensional dilatino currents
\begin{eqnarray}
&&\bar{\lambda}_{\cal {X}}^-\gamma^{\mu}\lambda_i^- = \nonumber \\
&& e^{i(P_i-P_{\cal {X}}) \cdot x} \frac{c_i' c_{\cal {X}}' M^{1/2}_{\cal {\chi}}}{z_0^{\tilde{m}_i+1}} z^{\tilde{m}_i+9/2} \bar{u}_{\sigma {\cal {X}}} \left(  J_{\tilde{m}_{\cal {X}}-\frac{1}{2}}(M_{\cal {X}} z)\gamma^{\mu} P_+ +   J_{\tilde{m}_{\cal {X}}-\frac{1}{2}}(M_{\cal {X}} z)\gamma^{\mu}P_- \frac{M_i z}{2 \tilde{m}_i+1}  \right) u_{\sigma i} \, , \nonumber \\
&& \\
&&\bar{\lambda}_{\cal {X}}^-\gamma^{z}\lambda_i^-= \nonumber \\
&& e^{i(P_i-P_{\cal {X}}) \cdot x} \frac{c_i' c_{\cal {X}}' M^{1/2}_{\cal {\chi}}}{z_0^{\tilde{m}_i+1}} z^{\tilde{m}_i+9/2}\bar{u}_{\sigma {\cal {X}}} \left(  J_{\tilde{m}_{\cal {X}}-\frac{1}{2}}(M_{\cal {X}} z) \frac{M_i z}{2 \tilde{m}_i+1}  (-P_-) +  J_{\tilde{m}_{\cal {X}}+\frac{1}{2}}(M_{\cal {X}} z)   P_+\right)u_{\sigma i} \, .  \nonumber \\
\end{eqnarray}
Next, we solve the $I_1$ and $I_2$ integrals, and introduce the twist defined as $\tau_i=\tilde{m_i}+3/2$. Then, we obtain the following matrix elements:
\begin{eqnarray}
I_1&=& i \frac{1}{2\pi} \left(\frac{M_{\cal {X}}}{z_0}\right)^{\frac{1}{2}} \left( \int d^4x e^{i(q+P_i-P_{\cal {X}}) \cdot x}\right) \times \nonumber \\
&& \left( 2 \pi c_i' c_{\cal {X}}' z_0^{-\tau_i+1}  \left(\int dz K_1(q z)z^{\tau_i} J_{\tau_{\cal {X}}-2}(M_{\cal {X}} z) \right)    q \bar{u}_{\sigma {\cal {X}}}\slashed{n} P_+ u_{\sigma i}  \right. \nonumber \\
&&  \left. +  2 \pi c_i' c_{\cal {X}}' z_0^{-\tau_i+1} \left(\int dz K_1(q z)z^{\tau_i+1} J_{\tau_{\cal {X}}-1}(M_{\cal {X}} z) \right)\frac{M_i}{2(\tau_i-1)}  q \bar{u}_{\sigma {\cal {X}}}\slashed{n} P_- u_{\sigma i} \right) \, . \nonumber \\
\end{eqnarray}
Similarly for the other integral
\begin{eqnarray}
I_2 &=& i \frac{1}{2\pi} \left(\frac{M_{\cal {X}}}{z_0}\right)^{\frac{1}{2}} \left(  \int d^4x e^{i(q+P_i-P_{\cal {X}}) \cdot x} \right) \times \nonumber \\
&& \left( 2 \pi c_i' c_{\cal {X}}' z_0^{-\tau_i+1} \left(- \int dz z^{\tau_i+1} J_{\tau_{\cal {X}}-2}(M_{\cal {X}} z) K_0(qz) \right)\left( i n \cdot q \right)\frac{M_i }{2(\tau_i-1)} \bar{u}_{\sigma {\cal {X}}} P_-u_{\sigma i} \right. \nonumber\\
&&   +\left. \pi c_i' c_{\cal {X}}' z_0^{-\tau_i+1}   \left( \int dz z^{\tau_i} K_0(qz) J_{\tau_{\cal {X}}-1}(M_{\cal {X}} z) \right)\left( i n \cdot q \right) \bar{u}_{\sigma {\cal {X}}}P_+u_{\sigma i}                   \right) \, .
\end{eqnarray}

The constants $c_n$ with $n=1, \dots 4$ are defined from certain Bessel function integrals in terms of $x$, $q$ and $M_i$
\begin{eqnarray}
c_1&=&c_0\int dz K_1(q z)z^{\tau_i} J_{\tau_{\cal {X}}-2}(M_{\cal {X}} z)  = c_0 2^{\tau-1} \Gamma(\tau) q^{-(\tau+1)} x^{\frac{\tau+2}{2}} (1-x)^{\frac{\tau-2}{2}}  \, , \\
c_2&=&c_0 \frac{M_i }{2(\tau_i-1)} \int dz K_1(q z)z^{\tau_i+1} J_{\tau_{\cal {X}}-1}(M_{\cal {X}} z) \nonumber \\
&=& c_0 2^{\tau-1} M_i \frac{\tau \Gamma(\tau)}{\tau-1} q^{-(\tau+2)} x^{\frac{\tau+3}{2}} (1-x)^{\frac{\tau-1}{2}} \, , \\
c_3&=&c_0 \frac{M_i }{2(\tau_i-1)} \int dz z^{\tau_i+1} J_{\tau_{\cal {X}}-2}(M_{\cal {X}} z) K_0(qz) \nonumber \\ 
&=& c_0 2^{\tau-1} M_i \frac{\Gamma(\tau)}{\tau-1} q^{-(\tau+2)} x^{\frac{\tau+2}{2}} (1-x)^{\frac{\tau-2}{2}} (x\tau-1) \, , \\
c_4&=&c_0 \int dz z^{\tau_i} K_0(qz) J_{\tau_{\cal {X}}-1}(M_{\cal {X}} z) = c_0 2^{\tau-1}\Gamma(\tau) q^{-(\tau+1)} x^{\frac{\tau+1}{2}} (1-x)^{\frac{\tau-1}{2}} \, .
\end{eqnarray}

\end{document}